\documentclass[12pt,a4paper]{JHEP3}
\pdfoutput=1
\usepackage[latin1]{inputenc}
\usepackage[english]{babel}
\usepackage{amsmath,amssymb}
\usepackage[pdftex]{graphicx}

\usepackage{cite}

\newcommand{\beq}{\begin{equation}}
\newcommand{\eeq}{\end{equation}}
\newcommand{\bea}{\begin{eqnarray}}
\newcommand{\eea}{\end{eqnarray}}
\newcommand{\bdm}{\begin{displaymath}}
\newcommand{\edm}{\end{displaymath}}
\newcommand{\eps}{\varepsilon}

\def\d{\mathrm{d}}
\def\as{\alpha_s}

\def\m{{\cal M}}

\def\ord{{\cal O}}
\def\eps{\varepsilon}

\def\dy{\Delta y}
\def\yb{\bar{y}}
\title{Jet vetoing at the LHC}
\author{Jeffrey Forshaw, James Keates and Simone Marzani \\School of Physics \& Astronomy, University of Manchester,\\Oxford Road, Manchester, M13 9PL, U.K.\\

\email{jeff.forshaw@manchester.ac.uk, \\ james.keates@cern.ch, \\ simone.marzani@manchester.ac.uk}}
\preprint{MAN/HEP/2009/16}

\keywords{QCD, Jets}

\abstract{We study the effect of a veto on additional jets in the rapidity region between a pair of high transverse momentum jets at the LHC. We aim to sum the most important logarithms in the ratio of the jet transverse momentum to the veto scale and to that end we attempt to assess the significance of the super-leading logarithms that appear at high orders in the perturbative expansion. We also compare our results to those of \textsc{Herwig++}, in an attempt to ascertain the accuracy of the angular ordered parton shower. We find that there are large corrections that arise for large enough jet transverse momenta as a consequence of Coulomb gluon exchanges.}

\begin{document}
\section{Introduction} \label{sec:intro}
Dijet production, with a veto on the emission of additional radiation in the inter-jet region has been widely studied in electron-proton collisions at HERA~\cite{zeus95,zeus06,H102} and in $p \bar{p}$ collisions at the Tevatron~\cite{D0gaps, cdfgaps}, and theoretical calculations have been compared to data~\cite{cox,Motyka:2001zh,Enberg:2001ev,Chevallier:2009cu,applebyHERA}. We shall refer generically to the ``gaps between jets'' process, although it is to be remembered that the veto scale can certainly be large and so a ``gap'' is simply a region of limited hadronic activity. Indeed we shall require the veto scale to be large in this paper in order to permit a calculation using perturbative QCD.

\begin{figure}
\begin{center}
\includegraphics[width=0.8\textwidth, clip]{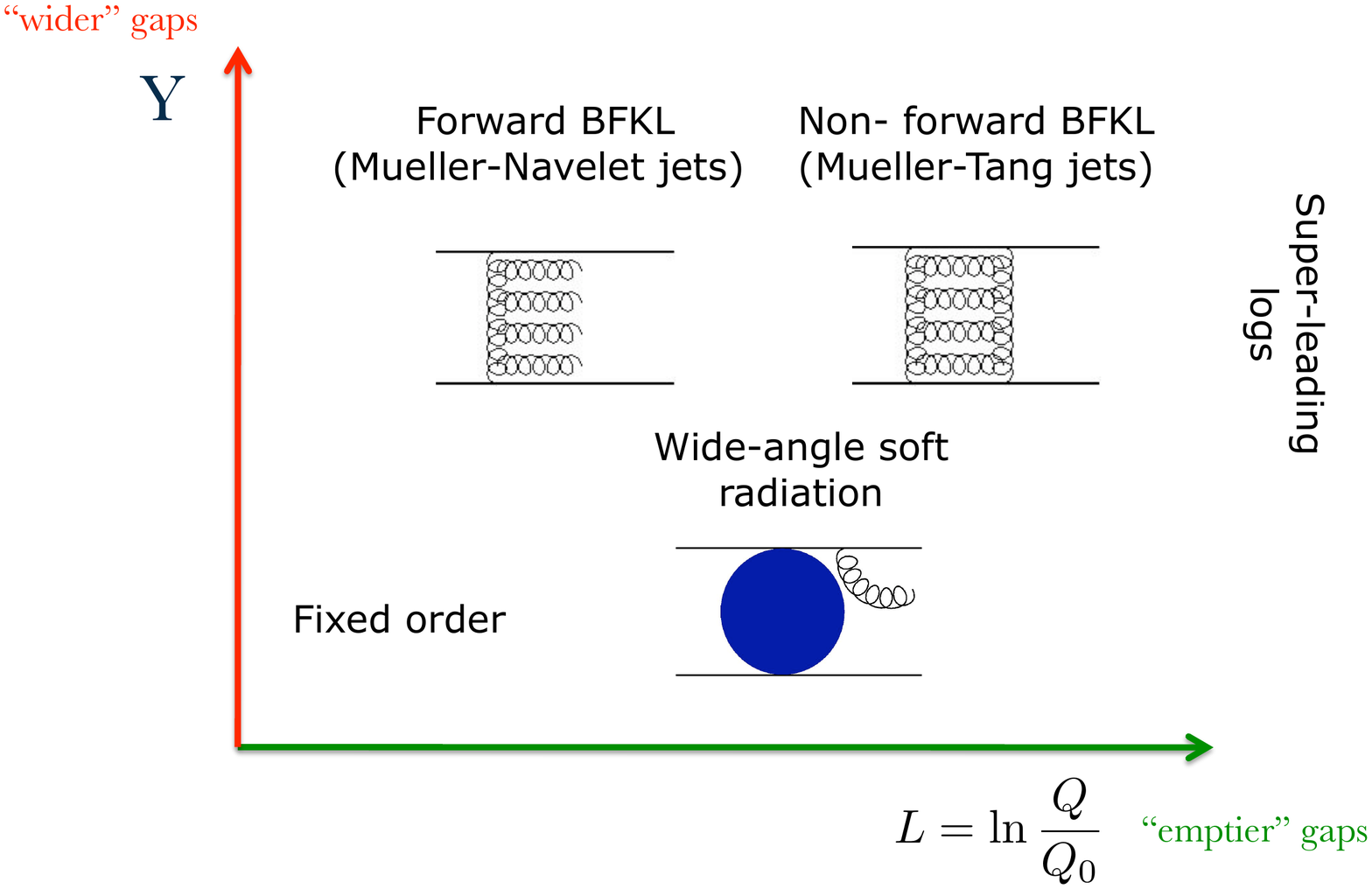}
\caption{A map of the $(L,Y)$ plane. }\label{fig:QCD}\end{center}
\end{figure}

Gaps between jets is an interesting process to study at the LHC. It is also a pure QCD process, hence the cross-section is large and studies can be performed with early data. It is interesting because it allows one to investigate a remarkably diverse range of QCD phenomena. These are summarised in Fig.~\ref{fig:QCD}, which maps out schematically the $(L,Y)$ plane, where $L=\ln (Q/Q_0)$ ($Q$ is the dijet $p_T$ and $Q_0$ the veto scale) and $Y$ is rapidity separation of the dijets. For instance, the limit of large rapidity separation corresponds to the limit of high partonic centre of mass energy and  BFKL effects are expected to become important~\cite{muellernavelet}.  On the other hand one can study the limit of emptier gaps, becoming more sensitive to wide-angle soft gluon radiation. Furthermore, if one wants to investigate both of these limits simultaneously, then  the non-forward BFKL equation enters the game~\cite{muellertang}.  In this paper we concentrate on wide-angle soft emissions.

Given a hard scattering process, we can study how it is modified by the addition of soft radiation. If the observable is inclusive enough, then we  have no effects because soft contributions cancel when real and virtual corrections are added together, as a result of the Bloch-Nordsieck theorem. However, if we restrict the real radiation to a corner of the phase space, as happens for the gap cross-section, we encounter a miscancellation and we are left with a logarithm of the ratio of the hard scale and veto scale.
The resummation of wide-angle soft radiation in the gaps between jets process was originally performed assuming that the real--virtual cancellation is perfect outside the gap, so that one need only consider virtual gluon corrections integrated over momenta for which real emissions are forbidden, i.e. over the ``in gap'' region of rapidity and with
$k_T$ above the veto scale $Q_0$~\cite{KOS,OS, Oderda}. However, it was later realised~\cite{DS} that one must also consider secondary emissions, i.e. real gluons emitted outside of the gap may emit back into the gap, and these emissions too are vetoed above $Q_0$. These give rise to a new tower of logarithms, formally as important as the primary emission corrections, known now as non-global logarithms. In order to obtain the complete tower of leading logarithms it is therefore not sufficient to consider only the virtual corrections to the $2 \to 2$ scattering process. To capture the non-global logarithms, one must also consider all $2 \to n $ processes, i.e. \mbox{$n-2$ out-of-gap gluons}, dressed with ``in-gap'' virtual corrections.  The colour structure quickly becomes intractable and, to date, corrections have been calculated numerically to all orders only in the large $N_c$ limit~\cite{DS,appleby2}. In~\cite{nonlinear} it was shown that the resummation of non-global logarithms in the large $N_c$ limit can be obtained by solving a non-linear evolution equation, which has the same form as the BK equation in small-$x$ physics \cite{Balitsky,Kovchegov} (see also~\cite{marchesinimueller}). This remarkable observation has stimulated further investigation into a correspondence between soft gluon emission and high energy (Regge) scattering~\cite{weigert,hatta1,hatta2}.
 
A different approach was taken in~\cite{SLL1}, where the specific case of only one gluon emitted outside the gap, dressed to all orders with virtual gluons but keeping the full $N_c$ structure, was considered. The surprising result of that calculation was the possible discovery of a new class of logarithms. These logarithms appear first at order $\alpha_s^4$ relative to the hard process and are formally more important than the primary emission logarithms at this order. These ``super-leading'' logarithms (SLL) are however sub-leading in colour, which is perhaps the reason why they were not identified in previous studies. Their origin can be traced to a failure of the DGLAP ``plus-prescription'', when the out-of-gap gluon becomes collinear to one of the incoming partons. Real and virtual contributions do not cancel as one would expect and one is left with an extra logarithm.  This miscancellation is caused by the imaginary part of loop integrals, induced by Coulomb gluons.  More recently the calculation has been repeated in a colour basis independent formalism~\cite{SLLind}. Such an approach allows one to consider, in principle, any number of gluons outside the gap. However, in order to perform an actual calculation, one must fix a colour basis and its increasing dimensionality hampers the computation. In this paper we limit ourselves to the resummation of SLL contributions coming only from the emission of only one gluon outside the gap. 
 
Accurate studies of these effects are important also in relation to other processes, in particular the production of a Higgs boson in association with two jets. It is well known that this process can occur via gluon-gluon fusion and weak-boson fusion (WBF). QCD radiation in the inter-jet region is clearly  different in the two cases and, in order to enhance the WBF channel, one can put a cut on emission between the jets \cite{Barger:1994zq,Kauer:2000hi}. This situation is very closely related to gaps between jets since the Higgs carries no colour charge. To date, the impact of a central jet veto has been explored in fixed-order calculations \cite{DelDuca:2004wt,Figy:2007kv,Arnold:2008rz}, using parton shower Monte Carlos \cite{DelDuca:2006hk,Nikitenko:2007it}, via soft gluon resummation \cite{softgluonshiggs} and in the framework of high energy factorisation \cite{Andersen:2008gc}. Control over the effects of QCD radiation in $Hjj$ is mandatory if we are accurately to extract the coupling of the Higgs to the weak bosons in this channel \cite{Zeppenfeld:2000td,Plehn:2001nj,Duhrssen:2004cv,Hankele:2006ma}. 

The remainder of this paper is organised as follows. In Section~\ref{sec:xsec} we define the observable and the kinematics. In Section~\ref{sec:zero} we reconsider the known case and sum up the subset of leading logarithms that arise as a result of imposing a veto on primary emissions into the gap. We refer to this as the ``zero gluons outside the gap'' cross-section and we study the effect of the resummation on the hadronic cross-section at the LHC. We compare our result to \textsc{Herwig++} and are able to assess the role of the Coulomb gluon corrections, which are not present in the angular ordered parton shower approach. In Section~\ref{sec:one} we perform the resummation of the SLL contributions that come from matrix elements in which there is one gluon emitted outside of the gap and we speculate on the size of the missing contributions from the emission of greater than one out-of-gap gluon. We finish with a look to the future.

  \section{The gaps-between-jets cross-section} \label{sec:xsec}
We are interested in dijet production in hadron-hadron collisions:
$$
h_1+h_2 \to jj +X\,,
$$
where we veto the emission of a third jet  with transverse momentum bigger than $Q_0 $ in the rapidity region between the two jets.
At leading order in perturbation theory we have to consider all the possible $2 \to 2$ parton scattering processes. The kinematics of such sub-processes can be expressed as follows:
\bea \label{partonkin}
p_a &=&  x_a \frac{\sqrt{S}}{2}(1,0,0,1) \nonumber \\
p_b &=&  x_b \frac{\sqrt{S}}{2}(1,0,0,-1) \nonumber \\
p_c &=& Q (\cosh y_c, \cos \varphi,\sin \varphi, \sinh y_c) \nonumber \\
p_d &=& Q (\cosh y_d, -\cos \varphi,-\sin \varphi, \sinh y_d)\,,
\eea
where
\beq
x_{a,b} = \frac{\sqrt{\rho}}{2} \left(e^{ \pm y_c}+ e^{\pm y_d} \right)\,, \quad \rho = \frac{4 Q^2}{S}\,.
\eeq
The triple differential cross-section is
\bea \label{trpldiff}
\frac{\d^3 \sigma}{\d y_c \d y_d \, \d Q^2} &=& \frac{g^4}{16 \pi S^2}  \sum_{a,b,c,d}  \frac{f_a(x_a,Q^2)}{x_a}\frac{f_b(x_b,Q^2)}{x_b} \nonumber \\ &&
|\m_{abcd}|^2   \frac{1}{1+\delta_{ab}}\frac{1}{1+\delta_{cd}}\,, \nonumber \\
\eea
where $|\m_{abcd}|^2$ is the colour-averaged matrix element squared and $f_{a,b}$ are the usual parton distribution functions for flavours $a$ and $b$. Notice that the coupling constant has been taken out of the matrix element for future convenience. 
Since we are interested in the gap cross-section, it is useful to express the kinematics in terms of
\bea
\bar{y} &=& y_c+y_d\,, \nonumber \\
\Delta y &=& y_c-y_d
\eea
and define 
\bea \label{zomega}
\omega &=& x_a/x_b= e^{\yb}\,, \nonumber \\
z &=&x_a x_b = \rho \cosh^2 \frac{\dy}{2}\,. 
\eea
We choose to define the kinematics for the generic process, 
\beq
a(p_a,i) \; b(p_b,j) \to c(p_c,k) \; d(p_d,l)\,,
\eeq 
such that parton $c$ is always left-moving, while $d$ is always right-moving; $i,j,k,l$ are colour indices. Hence we have that $y_c > y_d$, which means  $\dy > 0$, or, equivalently:
\bea \label{firsthalf}
0 \le &-t& \le \frac{s}{2}\,, \nonumber \\
\frac{s}{2} \le &-u& \le s\,, 
\eea
where the Mandelstam variables are defined by
\bea \label{studef}
s &=& (p_a+p_b)^2\,, \nonumber \\
t &=& (p_a-p_c)^2= -\frac{s}{2}\left(1- \tanh \frac{\dy}{2} \right)\,, \nonumber \\
u &=& (p_a-p_d)^2=-\frac{s}{2}\left(1+ \tanh \frac{\dy}{2} \right)\,.
\eea
At the Born level, the missing half of the phase space is obtained by exchanging \mbox{$t$ and $u$}. However, when resummation effects are taken into account, this  issue is more complicated and one must consider different anomalous dimension matrices, as explained in Section~\ref{sec:zero}.

In order to obtain the gaps-between-jets cross-section, we need to integrate Eq.~(\ref{trpldiff}) over the total rapidity of the parton system, $\yb$. Fortunately the squared matrix element does not depend upon $\yb$. This is trivial at LO and it remains true after the inclusion of the resummation effects. 
The gap is defined in terms of the distance in rapidity between the final-state partons, $\Delta y$, and the jet radius $R$:
\beq \label{gapdef}
Y = \Delta y - 2 R\,.
\eeq
The double differential cross-section is thus
\beq \label{gapsxsec}
Q^2\frac{\d^2 \sigma}{\d \dy\, \d Q^2} = \frac{\rho \pi \as^2}{4 S}  \sum_{a,b,c,d}  \mathcal{L}_{ab}(\dy,Q^2)|\m_{abcd}|^2\frac{1}{1+\delta_{ab}}\frac{1}{1+\delta_{cd}}
\eeq
and we have introduced the differential parton luminosity:
\beq  \label{difflum}
  \mathcal{L}_{ab}(\dy,Q^2) =\frac{1}{2 z} \int_{-\yb^+}^{\yb^+} d \yb \, 
 f_a(\sqrt{z} e^{\yb/2},Q^2) f_b(\sqrt{z} e^{-\yb/2},Q^2) \,,
\eeq
where the value of $z$ is determined by Eq.~(\ref{zomega}).  The integration limits are
\beq
\pm \yb^+ =\pm  {\rm min} \left(\ln \frac{1}{z},  \yb_{\rm cut}\right),
\eeq
where  the value $\yb_{\rm cut} = 2 \eta -Y - 2 R$ is obtained by requiring that both jets are within the calorimeter acceptance $\eta$. 
Since for both ATLAS and CMS the edge of the hadronic calorimeter is at $4.9$ units of rapidity, a sensible choice is
\beq
\eta = 4.5\: \quad {\rm for} \quad R=0.4\,.
\eeq
We also choose $Q_0= 20$~GeV as the veto scale. This is a fairly conservative choice in order to avoid contamination from the underlying event. It could be lowered once those effects are better understood.

The relevant flavour combinations of the parton distribution functions are
\bea \label{flavlum}
\mathcal{Q}^{(1)}(z,Q^2)&=& \frac{1}{2 z} \int_{-\yb^+}^{\yb^+}  d \yb 
\sum_{a}\left[ q_a(\sqrt{z} e^{\yb/2},Q^2) q_a(\sqrt{z} e^{-\yb/2},Q^2) \right. \nonumber \\ 
&&+ \left. \bar{q}_a(\sqrt{z} e^{\yb/2},Q^2) \bar{q}_a(\sqrt{z} e^{-\yb/2},Q^2)  \right] \,, \nonumber \\
\mathcal{Q}^{(2)}(z,Q^2)&=& \frac{1}{2 z} \int_{-\yb^+}^{\yb^+}  d \yb 
\sum_{a \neq b}\left[ q_a(\sqrt{z} e^{\yb/2},Q^2) q_b(\sqrt{z} e^{-\yb/2},Q^2) \right. \nonumber \\ 
&&+ \left. \bar{q}_a(\sqrt{z} e^{\yb/2},Q^2) \bar{q}_b(\sqrt{z} e^{-\yb/2},Q^2)  \right] \,,\nonumber \\
\mathcal{Q}^{(3)}(z,Q^2)&=& \frac{1}{2 z} \int_{-\yb^+}^{\yb^+}  d \yb 
\sum_{a}\left[ q_a(\sqrt{z} e^{\yb/2},Q^2) \bar{q}_a(\sqrt{z} e^{-\yb/2},Q^2) \right. \nonumber \\
&&+\left. \bar{q}_a(\sqrt{z} e^{\yb/2},Q^2) q_a(\sqrt{z} e^{-\yb/2},Q^2) \right] \,,\nonumber \\
\mathcal{Q}^{(4)}(z,Q^2)&=& \frac{1}{2 z} \int_{-\yb^+}^{\yb^+}  d \yb 
\sum_{a \neq b}\left[ q_a(\sqrt{z} e^{\yb/2},Q^2) \bar{q}_b(\sqrt{z} e^{-\yb/2},Q^2) \right. \nonumber \\ 
&&+ \left. \bar{q}_a(\sqrt{z} e^{\yb/2},Q^2) q_b(\sqrt{z} e^{-\yb/2},Q^2)  \right] \,,\nonumber \\
\mathcal{S}(z,Q^2)&=& \frac{1}{2 z} \int_{-\yb^+}^{\yb^+}  d \yb 
\sum_{a}\Big[  \nonumber \\ 
&&  \left( q_a(\sqrt{z} e^{\yb/2},Q^2)+ \bar{q}_a(\sqrt{z} e^{\yb/2},Q^2) \right)
g(\sqrt{z} e^{-\yb/2},Q^2)   \nonumber \\ 
&& \left( q_a(\sqrt{z} e^{-\yb/2},Q^2)+ \bar{q}_a(\sqrt{z} e^{-\yb/2},Q^2) \right)
g(\sqrt{z} e^{\yb/2},Q^2)  \Big] \,,\nonumber \\ 
\mathcal{G}(z,Q^2)&=& \frac{1}{2 z} \int_{-\yb^+}^{\yb^+}  d \yb 
\left[ g(\sqrt{z} e^{\yb/2},Q^2) g(\sqrt{z}  e^{-\yb/2},Q^2) \right] \,.
\eea
The hadronic cross-section is then computed by collecting together the various partonic contributions, multiplied by the appropriate parton luminosity:
\bea\label{explgapsxsec}
Q^2\frac{\d^2 \sigma^{(i)}}{\d Y\, \d Q^2} &=& \frac{\rho \pi \as^2}{4S} \Bigg[ 
 |\widetilde{\m}_{qqqq}|^2 \mathcal{Q}^{(1)}
+ |\widetilde{\m}_{qq'qq'}|^2 \mathcal{Q}^{(2)} \nonumber \\
&+& \left( |\widetilde{\m}_{q\bar{q}q\bar{q}}|^2+ (n_f-1)|\widetilde{\m}_{q\bar{q}q'\bar{q}'}|^2+|\widetilde{\m}_{q\bar{q}gg}|^2 \right)\mathcal{Q}^{(3)}
\nonumber \\&+& |\widetilde{\m}_{q\bar{q}'q\bar{q}'}|^2\mathcal{Q}^{(4)}+|\widetilde{\m}_{qgqg}|^2\mathcal{S} \nonumber \\&+& \left(n_f|\widetilde{\m}_{ggq\bar{q}}|^2+|\widetilde{\m}_{gggg}|^2 \right) \mathcal{G} \Bigg]\,,
\eea
where we have collected together the contributions coming from the subprocesses with four quarks and four antiquarks.
Henceforth we write the cross-section with a superscript $(i)$, which denotes the number of gluons outside the gap.
The tilde reminds us that we are generally considering resummed squared matrix elements.  It is the aim of Section~\ref{sec:zero} to compute such matrix elements in the case of no out-of-gap gluons, i.e. $i=0$. We  consider the case of one gluon outside the gap in Section~\ref{sec:one}.
In the following we also omit the double derivative notation, but it is understood that we always consider the double differential cross-section.

\section{Zero gluons outside the gap} \label{sec:zero}

The resummation of the global logarithms is achieved by considering the original four-parton matrix element, dressed by virtual gluons ``in the gap'', with transverse momenta above $Q_0$. No out-of-gap gluons are included. The formalism is well known and it was first applied to gaps-between-jets in~\cite{OS}. Here we recap the result and also address, in some detail, the issue of the $t-u$ symmetrisation, because we have found it to be less than clearly explained in the existing literature. We then study the effects of resummation at the LHC and compare our results to \textsc{Herwig++}.

The colour basis independent formalism was introduced in~\cite{CCM} and more recently applied to the gaps-between-jets cross-section in~\cite{SLLind}.
In this framework the resummation of global logarithms is achieved by considering
\beq \label{resummedpartonicxsec1}
|\m|^2 = \frac{1}{V_c} \langle m_0 | e^{- \xi \mathbf{\Gamma}^{\dagger}}e^{- \xi \mathbf{\Gamma}} |m_0 \rangle\,,
\eeq
where
\beq \label{rc}
\xi  =\xi(Q_0,Q)=\frac{2}{\pi} \int_{Q_0}^{Q} \frac{d k_T}{k_T} \as(k_T)
\eeq
and $V_c$ is an averaging factor for initial state colour.
The vector $|m_0 \rangle$ represents the Born amplitude and the operator $\mathbf{\Gamma}$ is the soft anomalous dimension:
\beq \label{gammaoperator}
\mathbf{\Gamma} = \frac{1}{2}Y \mathbf{t}_t^2+ i \pi \mathbf{t}_a\cdot \mathbf{t}_b +\frac{1}{4}\rho_{\rm jet}(Y, |\dy|)(\mathbf{t}_c^2+\mathbf{t}_d^2)\,, 
\eeq
where $\mathbf{t}_i$ is the colour charge of parton $i$ and
\beq \label{jetfunc}
\rho_{\rm jet}(Y,\dy)= \ln \frac{\sinh\left(\dy/2+ Y/2 \right)}{\sinh\left(\dy/2- Y/2 \right)}-Y\,.
\eeq
The operator $\mathbf{t}_i^a$ describes the mapping from an $m$ dimensional vector space onto an $m+1$ dimensional one, as a consequence of emitting a soft gluon with colour $a$:
\beq \label{tdef}
|m+1 \rangle= g \sum_i \frac{p_i \cdot \eps^*}{p_i \cdot k}\mathbf{t}_i^a|m \rangle\,.
\eeq
The operator $\mathbf{t}_t^2$ represents the colour exchanged in the $t$-channel, i.e. $\mathbf{t}_t=\mathbf{t}_a+\mathbf{t}_c=\mathbf{t}_b+\mathbf{t}_d$. Some care is needed however since the flavour of partons $c$ and $d$ changes under the $t \leftrightarrow u$ symmetrization. If we use the notation $12 \to 34$ to label the parton flavours then 
\beq \label{t13}
\mathbf{t}_t^2= (\mathbf{t}_1+\mathbf{t}_3)^2=(\mathbf{t}_2+\mathbf{t}_4)^2\, 
\eeq 
in the case that parton 3 is identified with parton $c$, and
\beq \label{t14}
\mathbf{t}_t^2= (\mathbf{t}_1+\mathbf{t}_4)^2=(\mathbf{t}_2+\mathbf{t}_3)^2\,
\eeq 
in the case that parton 4 is identified with parton $c$.

In order to compute cross-sections, one must fix a basis in the vector space spanned by $|m \rangle$ and consequently a representation for the colour operators has to be computed. We choose to represent the $4$-parton amplitude by $\m_{i,j,k,l}$, where the $i,j\,(k,l)$ are the colour indices of the incoming (outgoing) partons. 
In the case of the radiating parton being an outgoing quark or an incoming antiquark, the colour operators are represented by $t^a_{ij}$, the generators in the fundamental representation. In the case of an incoming gluon, they are $T^a_{bc} = -i f^a_{bc}$ and the sign reverses in the case of an incoming quark or an outgoing antiquark or gluon. We always work in an orthonormal basis, expecting symmetric results for the anomalous dimension matrices~\cite{EVqqqqg,ansym}. The expression in Eq.~(\ref{resummedpartonicxsec1}) then becomes
\beq \label{resummedpartonicxsec2}
|\m|^2 = \frac{1}{V_c}{\rm tr}  \left[\m_0^{\dagger} e^{- \xi \Gamma^{\dagger}}e^{- \xi \Gamma} \m_0 \right]=  {\rm tr} \left( HS\right)\,,
\eeq
where the hard and soft matrices, $H = \m_0 \m_0^{\dagger}/V_c$ and $S=e^{- \xi \Gamma^{\dagger}}e^{- \xi \Gamma}$ have been introduced. In the following sub-sections we analyse in detail the various channels that contribute to the cross-section. For each sub-process we fix a basis and compute the hard scattering matrix at the Born level ($H$) and the soft anomalous dimension $\Gamma$.  We usually fix $C_A= N_c=3$ and $C_F = \frac{N_c^2-1}{2 N_c}= \frac{4}{3}$.

\subsection{Anomalous dimension matrices}
We start by considering the scattering of non-identical quarks, $$qq' \to qq',$$ in the basis:
\bea \label{qqqqbasis}
c_1 &= & \frac{1}{3}\delta_{ik} \delta_{jl}\,, \nonumber \\
c_2 &=& \frac{1}{2\sqrt{2}} \left(\delta_{il} \delta_{jk}-\frac{1}{3}\delta_{ik} \delta_{jl} \right).
\eea
The hard scattering matrix is given by
\beq \label{qqpqqpH}
H(t,u)= \frac{4}{9}\left(\begin{array}{cc}
0 & 0\\
  0 &  \frac{u^2+s^2}{t^2}
 \end{array}
   \right)\,.
\eeq
The anomalous dimension matrix computed by projecting Eq.~(\ref{gammaoperator}) onto the basis specified by Eq.~(\ref{qqqqbasis}) is
\beq \label{qqqqgammat}
\Gamma_{13} = \left(\begin{array}{cc}
 \frac{2}{3} \rho_{\rm jet} & \frac{i \pi \sqrt{2}}{3} \\
  \frac{i \pi \sqrt{2}}{3} & \frac{2}{3} \rho_{\rm jet} + \frac{3}{2} Y- \frac{i \pi}{3}
\end{array}
   \right)\,,
\eeq
where the subscript $13$ indicates that we have considered 
$\mathbf{t}_t^2= (\mathbf{t}_1+\mathbf{t}_3)^2$ (in which case the basis is a $t$-channel singlet-octet basis).  

The anomalous dimension which describes the other half of the phase-space is given by
\beq \label{qqqqgammau}
\Gamma_{14} = \left(\begin{array}{cc}
 \frac{2}{3} \rho_{\rm jet}+\frac{4}{3} Y & \frac{i \pi \sqrt{2}}{3}-  \frac{\sqrt{2}}{3} Y \\
  \frac{i \pi \sqrt{2}}{3} -\frac{\sqrt{2}}{3}Y & \frac{2}{3} \rho_{\rm jet} + \frac{1}{6} Y- \frac{i \pi}{3}
\end{array}
   \right).
\eeq
The basis specified in Eq.~(\ref{qqqqbasis}) is then a $u$-channel basis.
The same matrices describe the subprocess with four antiquarks.
The resummed squared matrix element for non-identical quarks (or antiquarks) is then
\beq\label{resumqq1}
|\widetilde{\m}_{qq'qq'}|=  {\rm tr}\left[ H(t,u)S_{13}+H(u,t)S_{14} \right]\,.
\eeq

In the case of identical quark scattering  $$ qq \to qq \,$$ the hard scattering matrix is 
\beq \label{qqqqH}
H(t,u) = \frac{4}{9}\left(\begin{array}{cc}
\frac{8}{9} \chi_1 & \frac{2 \sqrt{2}}{9} \chi_2 \\
  \frac{2 \sqrt{2}}{9} \chi_2 &  \chi_3
 \end{array}
   \right)\,,
\eeq
with
\bea
\chi_1 & = & \frac{t^2+s^2}{u^2}\,, \nonumber \\
\chi_2 & = & 3 \frac{s^2}{u t}-\frac{t^2+s^2}{u^2}\,, \nonumber \\
\chi_3 & = & \frac{u^2+s^2}{t^2}+\frac{1}{9}\frac{t^2+s^2}{u^2}-\frac{2}{3} \frac{s^2}{u t}\,.
\eea
The resummed squared matrix element for the case of four identical quarks (or antiquarks) is 
\beq\label{resumqq}
|\widetilde{\m}_{qqqq}|=  \frac{1}{2} {\rm tr}\left[ H(t,u)S_{13}+H(u,t)S_{14}\right]=
{\rm tr}\left[H(t,u)S_{13}\right] \,,
\eeq
as expected for identical final state particles.

We now  consider the scattering processes involving two quarks and two antiquarks,
$$   q\bar{q} \to q\bar{q},  $$
for which we choose the basis:
\bea \label{qbarqqbarbasis}
c_1 &= & \frac{1}{3}\delta_{ik} \delta_{jl}\,, \nonumber \\
c_2 &=& \frac{1}{2\sqrt{2}} \left(\delta_{ij} \delta_{kl}-\frac{1}{3}\delta_{ik} \delta_{jl} \right)\,.
\eea
The hard scattering matrix is
\beq \label{qbarqqbarH}
H(t,u) = \frac{4}{9}\left(\begin{array}{cc}
\frac{8}{9} \chi_1 & \frac{2 \sqrt{2}}{9} \chi_2 \\
  \frac{2 \sqrt{2}}{9} \chi_2 &  \chi_3
 \end{array}
   \right)\,,
\eeq
where
\bea
\chi_1 & = & \frac{t^2+u^2}{s^2}\,, \nonumber \\
\chi_2 & = & 3 \frac{u^2}{s t}-\frac{t^2+u^2}{s^2}\,, \nonumber \\
\chi_3 & = & \frac{u^2+s^2}{t^2}+\frac{1}{9}\frac{t^2+u^2}{s^2}-\frac{2}{3} \frac{u^2}{s t}\,.
\eea
 The two anomalous dimension matrices in the basis given by Eq.~(\ref{qbarqqbarbasis}) are
\beq \label{qbarqqbargammat}
\Gamma_{13} = \left(\begin{array}{cc}
 \frac{2}{3} \rho_{\rm jet} & \frac{-i \pi \sqrt{2}}{3} \\
  \frac{-i \pi \sqrt{2}}{3} & \frac{2}{3} \rho_{\rm jet} + \frac{3}{2} Y- \frac{7i \pi}{6}
\end{array}
   \right)\,
\eeq
and
\beq \label{qbarqqbargammau}
\Gamma_{14} = \left(\begin{array}{cc}
 \frac{2}{3} \rho_{\rm jet}+\frac{4}{3}Y & \frac{-i \pi \sqrt{2}}{3}+\frac{\sqrt{2}}{3} Y \\
  \frac{-i \pi \sqrt{2}}{3}+\frac{\sqrt{2}}{3} Y & \frac{2}{3} \rho_{\rm jet} +  Y- \frac{7i \pi}{6}
\end{array}
   \right)\,.
\eeq
The expression for the resumed matrix element is obtained by adding the two contributions:
\beq \label{resumqqbar}
|\widetilde{\m}_{q\bar{q}q\bar{q}}|^2 = {\rm tr} \left[ H(t,u)S_{13}+H(u,t)S_{14} \right]\,.
\eeq
For the unequal flavour process $q\bar{q}'\to q\bar{q}'$, one must drop the $s$-channel terms,  so that what is left is
\bea
\chi'_1 & = & 0\,, \nonumber \\
\chi'_2 & = & 0\,, \nonumber \\
\chi'_3 & = & \frac{u^2+s^2}{t^2}\,.
\eea
For $q\bar{q}\to q'\bar{q}'$ one drops the $t$-channel contributions, obtaining
\bea
\chi''_1 & = & \frac{t^2+u^2}{s^2}\,, \nonumber \\
\chi''_2 & = & -\frac{t^2+u^2}{s^2}\,, \nonumber \\
\chi''_3 & = & \frac{1}{9}\frac{t^2+u^2}{s^2}\,.
\eea

We now turn our attention to  quark-gluon scattering:
$$qg \to qg \quad {\rm and} \quad  \bar{q}g \to \bar{q}g\,.$$
A suitable basis is 
\bea \label{qgqgbasis}
c_1 &= & \frac{1}{\sqrt{24}}\,\delta_{ik} \delta_{jl}\,, \nonumber \\
c_2 &=& \sqrt{\frac{3}{20}}\, d_{jlc}t^c_{ki} \,, \nonumber \\
c_3 & =& \frac{1}{\sqrt{12}}\, i f_{jlc}t^c_{ki}\,,
\eea
where now $(j,l)$ are adjoint indices. In such a basis the hard scattering matrix is
\beq \label{qgqgH}
H(t,u) = \frac{1}{24}\left(
\begin{array}{lll}
 \frac{4 \text{$\chi_1 $}}{3} & \frac{2 \sqrt{10} \text{$\chi_1
   $}}{3} & 4 \sqrt{2} \text{$\chi_2 $} \\
 \frac{2 \sqrt{10} \text{$\chi_1 $}}{3} & \frac{10 \text{$\chi_1
   $}}{3} & 4 \sqrt{5} \text{$\chi_2 $} \\
 4 \sqrt{2} \text{$\chi_2 $} & 4 \sqrt{5} \text{$\chi_2 $} & 12
   \text{$\chi_3 $}
\end{array}
\right)\,,
\eeq
where
\bea
\chi_1 & = & 2-\frac{t^2}{s u}\,, \nonumber \\
\chi_2 & = & 1- \frac{1}{2}\frac{t^2}{s u}- \frac{u^2}{s t}-\frac{s}{t}\,, \nonumber \\
\chi_3 & = & 3-4 \frac{s u}{t^2}-\frac{1}{2}\frac{t^2}{s u}\,.
\eea
The anomalous dimension matrices turn out to be
\beq \label{qgqggamma}
\Gamma_{13}= \left(\begin{array}{ccc}
 \frac{13}{12} \rho_{\rm jet}+\frac{3 i \pi}{4}  & 0 &  \frac{\sqrt{2}i \pi}{2} \\
 0 & \frac{13}{12} \rho_{\rm jet}+\frac{3}{2}Y & \frac{\sqrt{5}i \pi}{4} \\
  \frac{\sqrt{2}i \pi}{2} &\frac{\sqrt{5}i \pi}{4} &\frac{13}{12} \rho_{\rm jet}+\frac{3}{2}Y
\end{array}
   \right)\,
\eeq
and
\beq \label{qgqggammau}
\Gamma_{14} = \left(\begin{array}{ccc}
 \frac{13}{12} \rho_{\rm jet}+\frac{3 i \pi}{4}+\frac{13}{6}Y  & 0 &  \frac{\sqrt{2}i \pi}{2}-\frac{\sqrt{2}}{2}Y \\
 0 & \frac{13}{12} \rho_{\rm jet}+\frac{17}{12}Y & \frac{\sqrt{5}i \pi}{4}-\frac{\sqrt{5}}{4}Y \\
  \frac{\sqrt{2}i \pi}{2}-\frac{\sqrt{2}}{2}Y &\frac{\sqrt{5}i \pi}{4}-\frac{\sqrt{5}}{4}Y &\frac{13}{12} \rho_{\rm jet}+\frac{17}{12}Y
\end{array}
   \right)\,.
\eeq
The hard scattering matrix and the anomalous dimensions for antiquark-gluon scattering have the same expressions as the ones stated above if one chooses a new basis, obtained from Eq.~(\ref{qgqgbasis}) by exchanging $i \leftrightarrow k$ and $j \leftrightarrow l$.
The resummed result is
\beq \label{resumqg}
|\widetilde{\m}_{qgqg}|^2 = {\rm tr} \left[ H(t,u)S_{13}+H(u,t)S_{14} \right]\,.
\eeq

For the process
$$gg \to q\bar{q}$$
 it is convenient to define an $s$-channel basis
\bea \label{ggqqbarbasis}
c_1 &= & \frac{1}{\sqrt{24}}\,\delta_{ij} \delta_{kl}\,, \nonumber \\
c_2 &=& \sqrt{\frac{3}{20}}\, d_{ijc}t^c_{kl} \,, \nonumber \\
c_3 & =& \frac{1}{\sqrt{12}}\, i f_{ijc}t^c_{kl}\,.
\eea
The hard scattering matrix is then given by
\beq \label{ggqqbarH}
H(t,u) = \frac{1}{V_c}\left(
\begin{array}{lll}
 \frac{4 \text{$\chi_1 $}}{3} & \frac{2 \sqrt{10} \text{$\chi_1
   $}}{3} & 2 \sqrt{2} \text{$\chi_2 $} \\
 \frac{2 \sqrt{10} \text{$\chi_1 $}}{3} & \frac{10 \text{$\chi_1
   $}}{3} & 2 \sqrt{5} \text{$\chi_2 $} \\
 2 \sqrt{2} \text{$\chi_2 $} & 2 \sqrt{5} \text{$\chi_2 $} & 6
   \text{$\chi_3 $}
\end{array}
\right)\,
\eeq
with  $V_c=(N_c^2-1)^2=64 $ and
the functions $\chi_i$ are
\bea \label{ggqbarchi}
\chi_1 & = & \frac{t}{u}+ \frac{u}{t}\,, \nonumber \\
\chi_2 & = & -2 \frac{u-t}{s}-\frac{t^2-u^2}{tu}\,, \nonumber \\
\chi_3 & = & \frac{t}{u}+ \frac{u}{t}-4\frac{t^2+u^2}{s^2}\,.
\eea
The anomalous dimensions are now
\beq \label{ggqbargamma}
\Gamma_{13} = \left(\begin{array}{ccc}
 \frac{2}{3} \rho_{\rm jet} +\frac{13}{6}Y-\frac{3 i \pi}{2} & 0 & - \frac{\sqrt{2}}{2} Y \\
 0 & \frac{2}{3} \rho_{\rm jet} +\frac{17}{12}Y & -\frac{\sqrt{5}}{4} Y \\
 - \frac{\sqrt{2}}{2} Y & -\frac{\sqrt{5}}{4}Y &  \frac{2}{3} \rho_{\rm jet} +\frac{17}{12}Y 
\end{array}
   \right)\,,
\eeq
\beq \label{ggqbargammamod}
\Gamma_{14} = \left(\begin{array}{ccc}
 \frac{2}{3} \rho_{\rm jet} +\frac{13}{6}Y-\frac{3 i \pi}{2} & 0 &  \frac{\sqrt{2}}{2} Y \\
 0 & \frac{2}{3} \rho_{\rm jet} +\frac{17}{12}Y & \frac{\sqrt{5}}{4} Y \\
  \frac{\sqrt{2}}{2} Y & \frac{\sqrt{5}}{4}Y &  \frac{2}{3} \rho_{\rm jet} +\frac{17}{12}Y 
\end{array}
   \right)\,.
\eeq
The resummed result is
\beq \label{resumggqq}
|\widetilde{\m}_{ggq \bar{q}}|^2 = {\rm tr} \left[ H(t,u)S_{13}+H(u,t)S_{14} \right]\,.
\eeq

The process
$$ q\bar{q} \to gg $$
is related to the previous one by time reversal. 
In the $s$-channel basis:
\bea \label{qbarggbasis}
c_1 &= & \frac{1}{\sqrt{24}}\,\delta_{ji} \delta_{kl}\,, \nonumber \\
c_2 &=& \sqrt{\frac{3}{20}}\, d_{lkc}t^c_{ji}  \,,\nonumber \\
c_3 & =& \frac{1}{\sqrt{12}}\, i f_{lkc}t^c_{ji}\,,
\eea
the hard scattering matrix is the same as Eq.~(\ref{ggqqbarH}) but with \mbox{$V_c=N_c^2=9$}. As far as the anomalous dimension is concerned the only difference with respect to Eq.~(\ref{ggqbargamma}) is in the coefficient in front of the jet function $\rho_{\rm jet}$:
\beq \label{qbargggamma}
\Gamma_{13} = \left(\begin{array}{ccc}
 \frac{3}{2} \rho_{\rm jet} +\frac{13}{6}Y-\frac{3 i \pi}{2} & 0 & - \frac{\sqrt{2}}{2} Y \\
 0 & \frac{3}{2} \rho_{\rm jet} +\frac{17}{12}Y & -\frac{\sqrt{5}}{4} Y \\
 - \frac{\sqrt{2}}{2} Y & -\frac{\sqrt{5}}{4}Y &  \frac{3}{2} \rho_{\rm jet} +\frac{17}{12}Y 
\end{array}
   \right)\,.
\eeq
Similarly 
\beq \label{qbargggammamod}
\Gamma_{14} = \left(\begin{array}{ccc}
 \frac{3}{2} \rho_{\rm jet} +\frac{13}{6}Y-\frac{3 i \pi}{2} & 0 &  \frac{\sqrt{2}}{2} Y \\
 0 & \frac{3}{2} \rho_{\rm jet} +\frac{17}{12}Y & \frac{\sqrt{5}}{4} Y \\
 \frac{\sqrt{2}}{2} Y & \frac{\sqrt{5}}{4}Y &  \frac{3}{2} \rho_{\rm jet} +\frac{17}{12}Y 
\end{array}
   \right)\,.
\eeq
Then the resummed result is 
\beq \label{resumqqgg}
|\widetilde{\m}_{q \bar{q}gg}|^2 = \frac{1}{2} {\rm tr} \left[ H(t,u)S_{13}+ H(u,t)S_{14}\right]= {\rm tr} \left[ H(t,u)S_{13}\right]\,.
\eeq
as expected for identical final-state partons. 

Finally we need to compute the four-gluon scattering process
$$gg \to gg\,.$$
We choose the following basis:
\bea \label{ggggbasis}
c_1 &= &  \frac{1}{8} \delta_{ik} \delta_{jl} \,, \nonumber \\
c_2 &=&  \frac{1}{2\sqrt{2}}\frac{3}{5} d_{ikc} d_{jlc}\,, \nonumber \\
c_3 &=&  \frac{1}{2\sqrt{2}}\frac{1}{3} f_{ikc} f_{jlc}\,, \nonumber \\
c_4 &=&  \frac{1}{2\sqrt{5}}\left[\frac{1}{2}\left(\delta_{ij} \delta_{kl}-\delta_{il} \delta_{jk}\right)-\frac{1}{3} f_{ikc} f_{jlc}\right]\,, \nonumber \\
c_5 &=&  \frac{1}{3\sqrt{3}}\left[\frac{1}{2}\left(\delta_{ij} \delta_{kl}+\delta_{il} \delta_{jk}\right)-\frac{1}{8}\delta_{ik} \delta_{jl}-\frac{3}{5} d_{ikc} d_{jlc}\right] \,.
\eea
The hard scattering matrix is
\beq \label{ggggH}
H(t,u)= \frac{1}{16}
\left(
\begin{array}{lllll}
 9 \text{$\chi_1 $} & 9 \sqrt{2} \text{$\chi_1 $} & 9 \sqrt{2}
   \text{$\chi_2 $} & 0 & -9 \sqrt{3} \text{$\chi_1 $} \\
 9 \sqrt{2} \text{$\chi_1 $} & 18 \text{$\chi_1 $} & 18
   \text{$\chi_2 $} & 0 & -9 \sqrt{6} \text{$\chi_1 $} \\
 9 \sqrt{2} \text{$\chi_2 $} & 18 \text{$\chi_2 $} & 8 \text{$\chi_3
   $} & 0 & -9 \sqrt{6} \text{$\chi_2 $} \\
 0 & 0 & 0 & 0 & 0 \\
 -9 \sqrt{3} \text{$\chi_1 $} & -9 \sqrt{6} \text{$\chi_1 $} & -9
   \sqrt{6} \text{$\chi_2 $} & 0 & 27 \text{$\chi_1 $}
\end{array}
\right)\,,
\eeq
where
\bea
\chi_1 & = & 1-\frac{t u}{s^2}-\frac{s t}{u^2}+\frac{t^2}{s u}\,, \nonumber \\
\chi_2 & = & \frac{s t}{u^2}-\frac{u t}{s^2}+ \frac{u^2}{s t}-\frac{s^2}{t u}\,, \nonumber \\
\chi_3 & = & \frac{27}{4}-9 \left(\frac{s u}{t^2}+\frac{1}{4}\frac{t u}{s^2}+\frac{1}{4} \frac{s t}{u^2}\right)
+ \frac{9}{2} \left(\frac{u^2}{s t}+\frac{s^2}{u t}-\frac{1}{2}\frac{t^2}{su} \right)
\,.
\eea
We find for the anomalous dimension matrix:
\bea \label{gggggamma}
\Gamma = \frac{3}{2}\rho_{\rm jet} \mathbb{I}_{5\times 5}+\left(\begin{array}{ccccc}
 \frac{3 i \pi}{4} &0 & - \frac{3\sqrt{2} i \pi}{4} &0 &0\\
 0 &\frac{3}{2}Y & -\frac{3 i \pi}{4}& -\frac{3 \sqrt{10} i \pi}{10} &0 \\
  - \frac{3\sqrt{2} i \pi}{4} & -\frac{3 i \pi}{4}& \frac{3}{2}Y & 0& -\frac{\sqrt{6} i \pi}{4}\\
  0 & -\frac{3 \sqrt{10} i \pi}{10} & 0 &3 Y-\frac{3 i \pi}{4} & -\frac{\sqrt{15} i \pi}{5} \\
  0 & 0& -\frac{\sqrt{6} i \pi}{4}& -\frac{\sqrt{15} i \pi}{5}& 4 Y-\frac{5 i \pi}{4}
\end{array}
   \right)\,.
\eea
We have already shown that in case of identical final state particles there is no need to compute $\Gamma_{14}$. The resummed squared matrix element is thus
\beq \label{resumgggg}
|\widetilde{\m}_{gggg}|^2 = {\rm tr} \left[ H(t,u)S_{13}\right]\,.
\eeq

\subsection{Phenomenology and comparison to HERWIG++}
In this section we compute the gaps-between-cross-section at the LHC, comparing the result obtained from the resummation of global logarithms to the one generated by a leading order parton shower Monte Carlo event generator, \textsc{Herwig++}~2.3.0 \cite{ThePEG,Bahr:2008pv,KLEISSCERN9808v3pp129,Gieseke:2003rz}. In both cases we use the the latest MSTW LO parton distribution functions \cite{mstw08}.

\begin{figure} [t]
\begin{center}
\includegraphics[width=0.48\textwidth, clip]{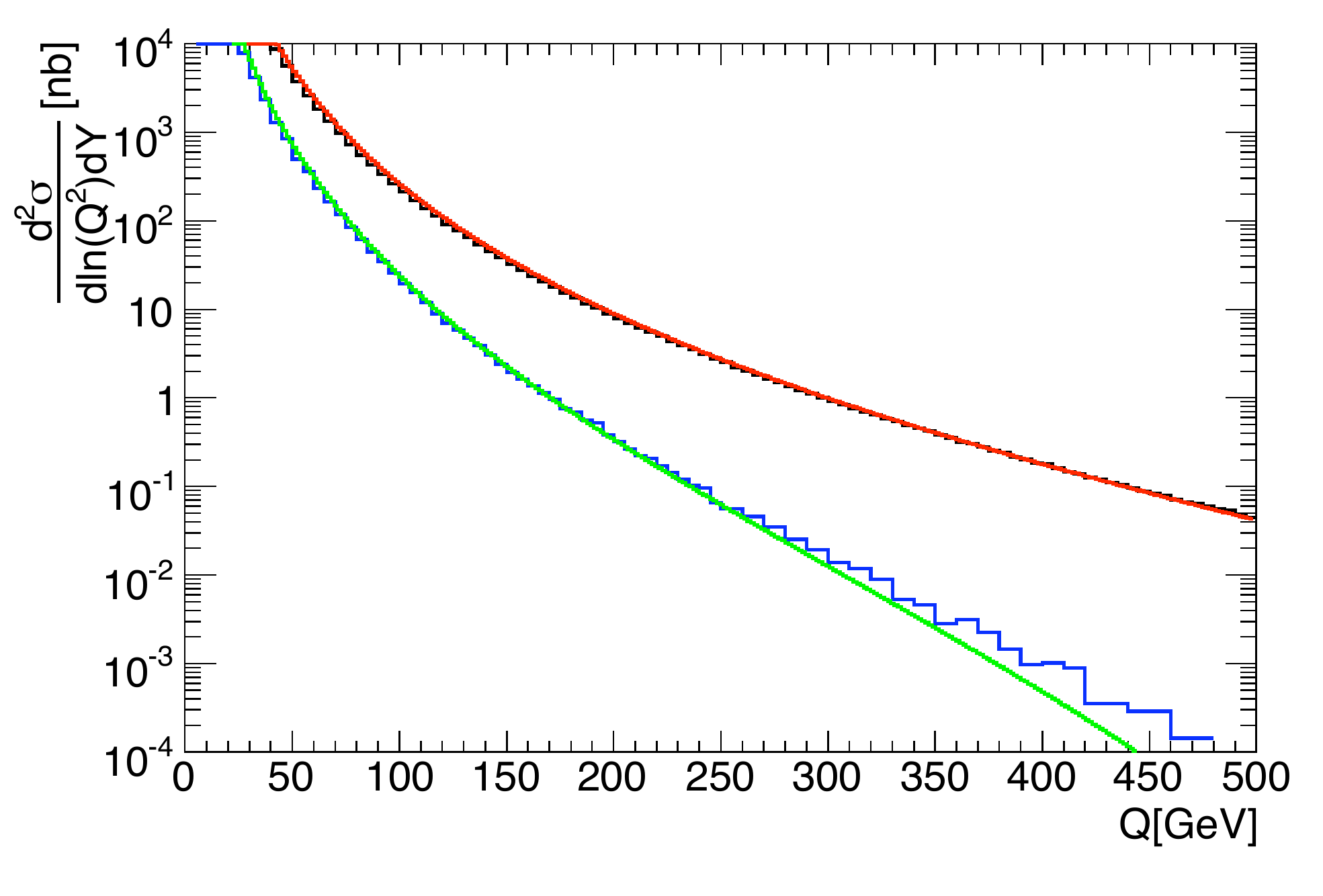}
\includegraphics[width=0.51\textwidth, clip]{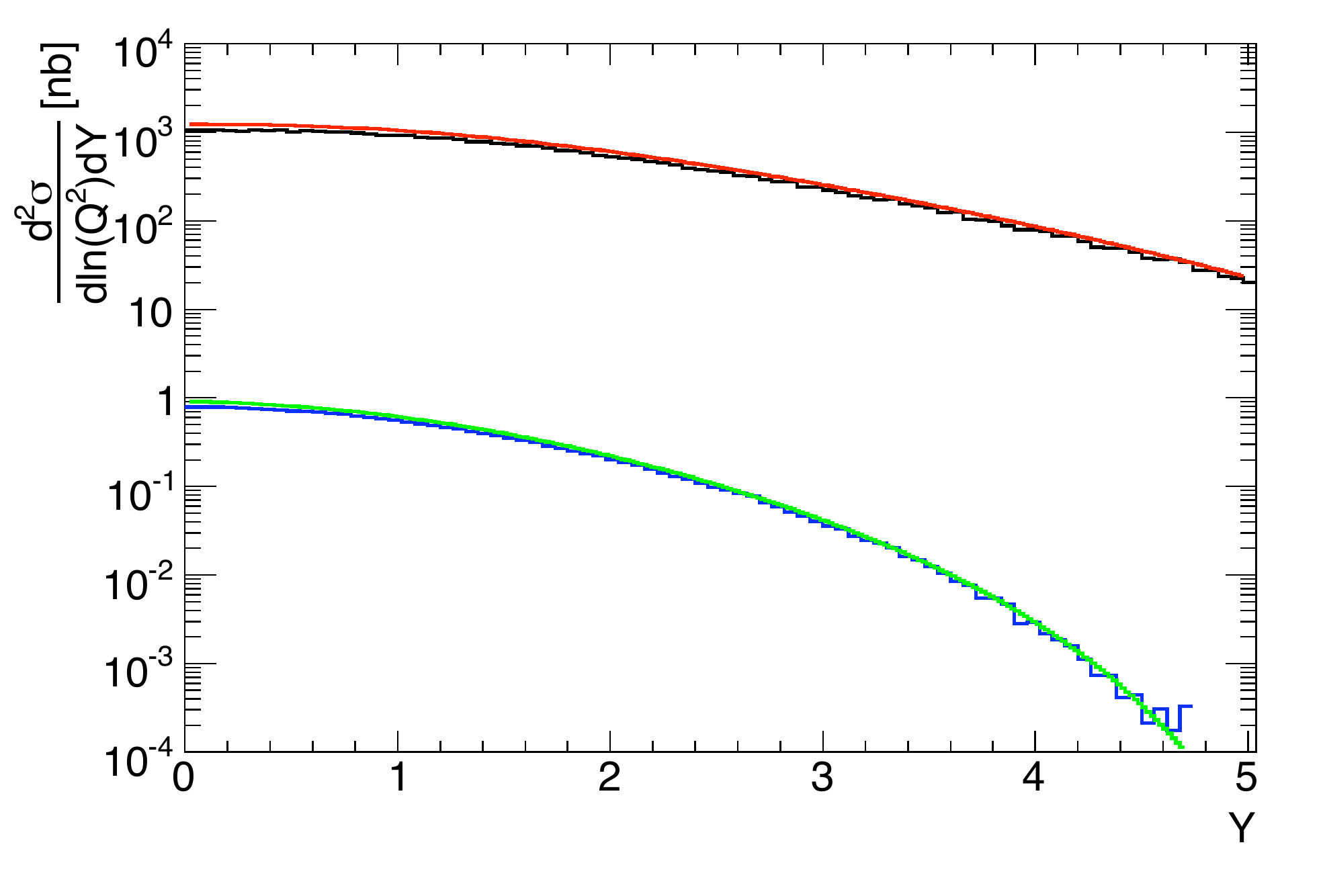}
\caption{The double differential Born cross-section (nb) computed using Eq.~(\ref{explgapsxsec}) without any resummation 
(solid lines), compared to the one obtained using \textsc{Herwig++} at the parton level (histograms) and without any parton showering. On the left the cross-section is plotted as a function of $Q$, for $Y=3$ (upper lines) and $Y=5$ (lower lines). On the right it is plotted  as a function of $Y$, for $Q=100$~GeV (upper lines) and $Q=500$~GeV (lower lines).} \label{fig:xsec1}
\end{center}
\end{figure}

As a cross-check, we first compare the result of the Born level calculation. The results are presented in
Fig.~\ref{fig:xsec1}. The double differential Born cross-section is computed using Eq.~(\ref{explgapsxsec}) without any resummation and it is compared to that obtained using \textsc{Herwig++} at the parton level but without any parton showering. The agreement is good, although there is a small (10\%) difference that arises because 
the computation of the cross-section from Eq.~(\ref{explgapsxsec}) uses the value of the strong coupling constant provided by the parton distribution function fit, whilst \textsc{Herwig++} uses its own one-loop coupling.  The leading order MSTW fit produces a larger value of $\as$ and this explains the difference.

\begin{figure} [t]
\begin{center}
\includegraphics[width=0.49\textwidth, clip]{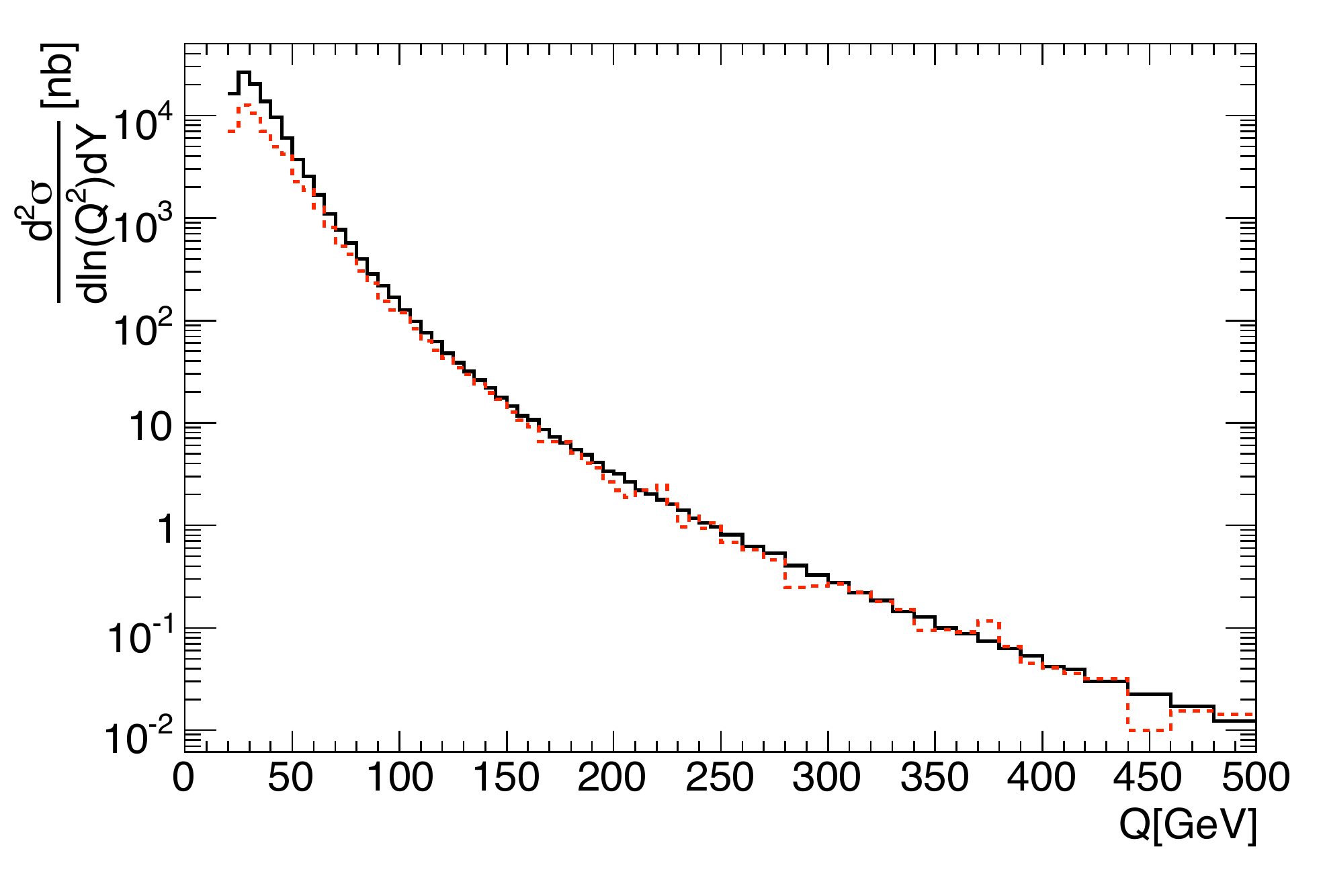}
\includegraphics[width=0.49\textwidth, clip]{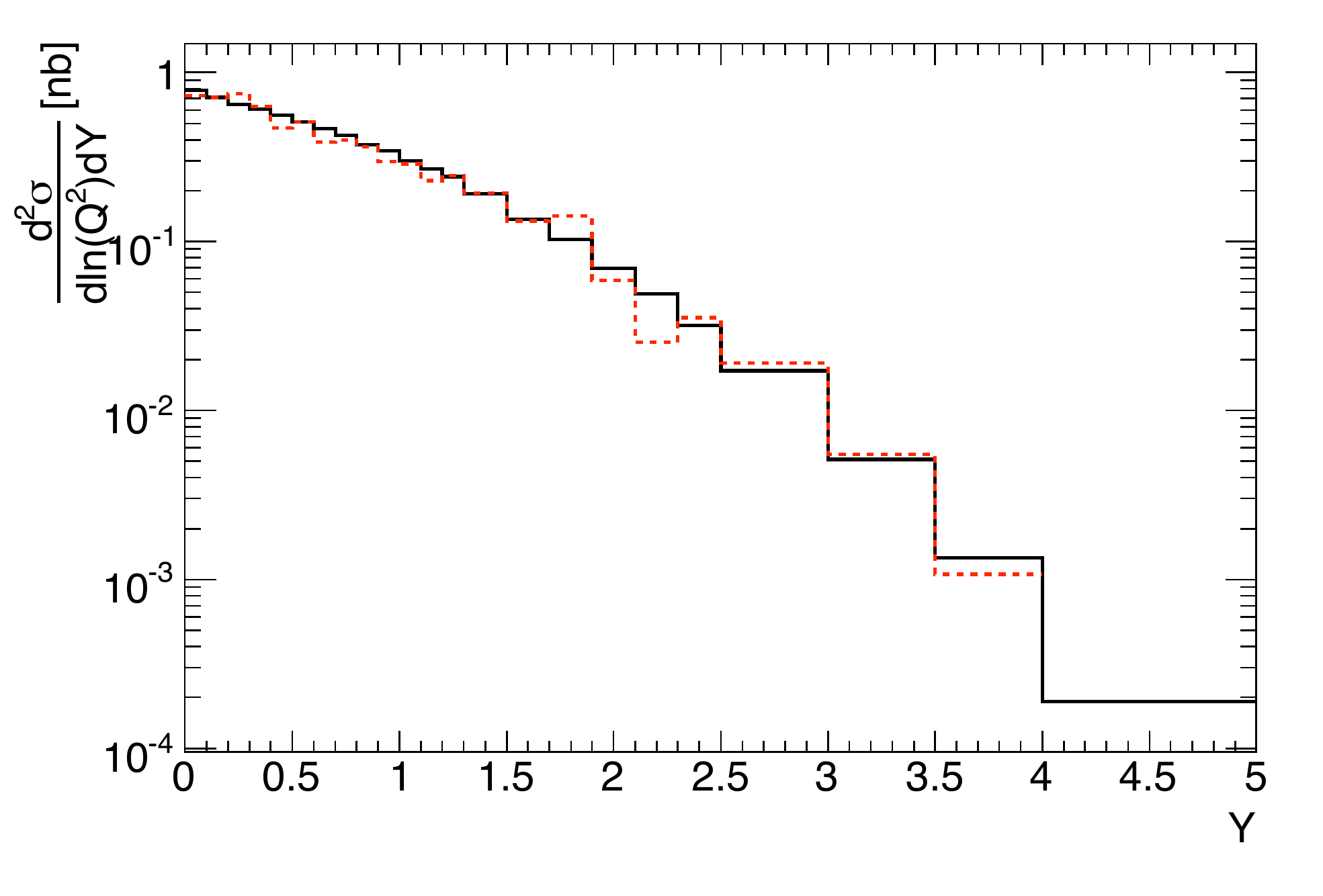}
\caption{The gap cross-section as a function of $Q$, for $Y=3$, and as a function of $Y$, for $Q=500$~GeV, at both parton shower level (solid) and after hadronisation (dotted).} \label{fig:hadron}
\end{center}
\end{figure}

We next present results after parton showering and jet finding.  
The jets are defined using the \textsc{Siscone}\cite{Salam:2007xv} algorithm, with a cone radius $R=0.4$ and an overlap parameter, $f=0.5$. The hardest two jets, which define the extremities of the gap, are required to have $|y_i|<4.5$ and $Q > 20$~GeV. We impose a veto such that there should be no third jet in-between the pair of hardest jets with $k_T$ above $20$~GeV. To assess the impact of hadronisation corrections, Fig.~\ref{fig:hadron} shows the gap cross-section as a function of $Q$, for $Y=3$, and as a function of $Y$, for $Q=500$~GeV, at both parton shower level and after hadronisation. The effects are modest and henceforth we always compare to the Monte Carlo results after parton showering but before hadronisation. We remark that corrections from the underlying event (which we have not included) ought also to be small due to our conservative choice of veto scale.

\begin{figure} [th]
\begin{center}
\includegraphics[width=0.49\textwidth, clip]{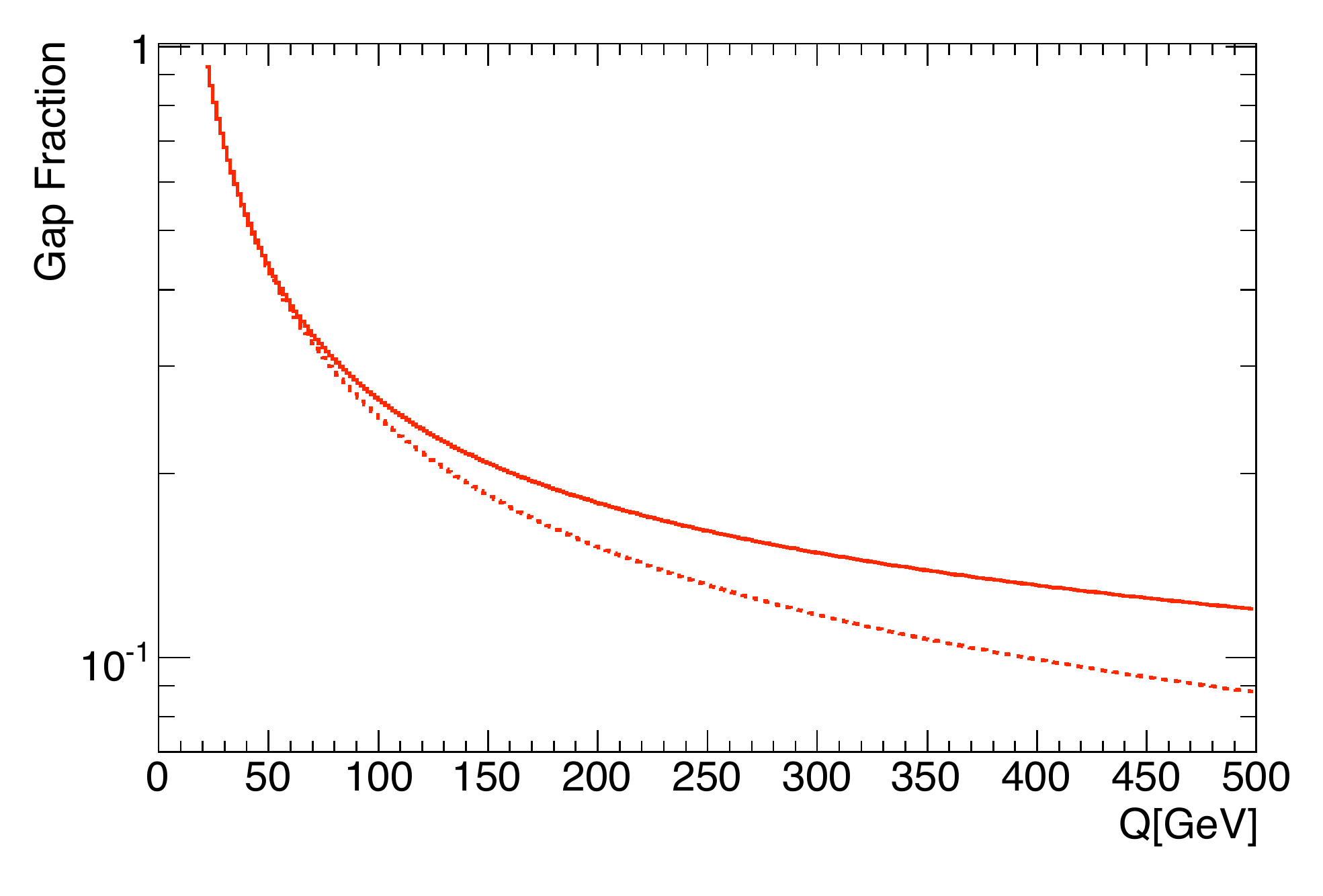}
\includegraphics[width=0.49\textwidth, clip]{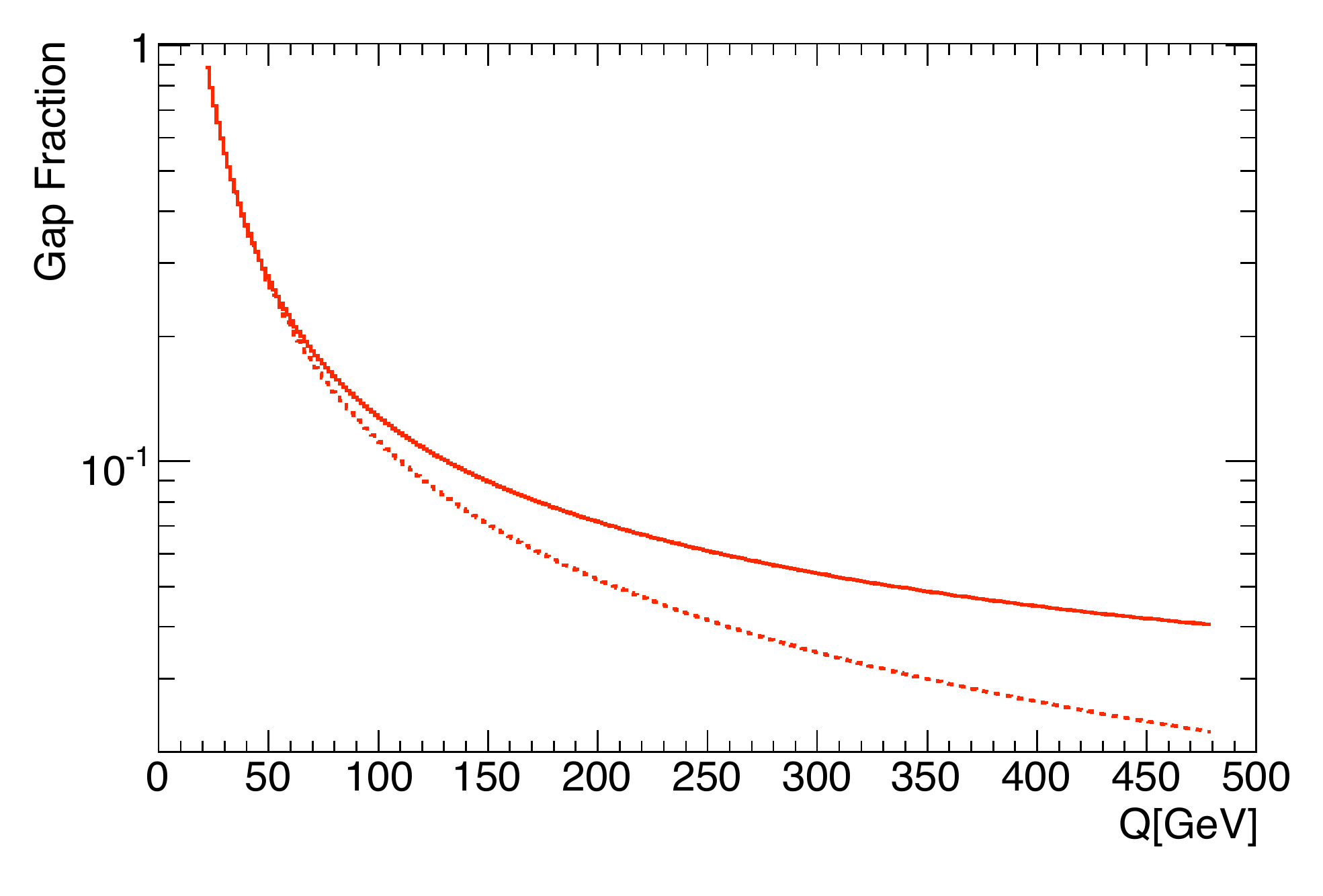}
\caption{Plot of the gap fraction as a function of $Q$, for $Y=3$ (left) and for $Y=5$ (right). The dashed lines are obtained by dropping the imaginary part of the evolution matrices.} \label{fig:kfact1}
\end{center}
\end{figure}

\begin{figure} [th]
\begin{center}
\includegraphics[width=0.49\textwidth, clip]{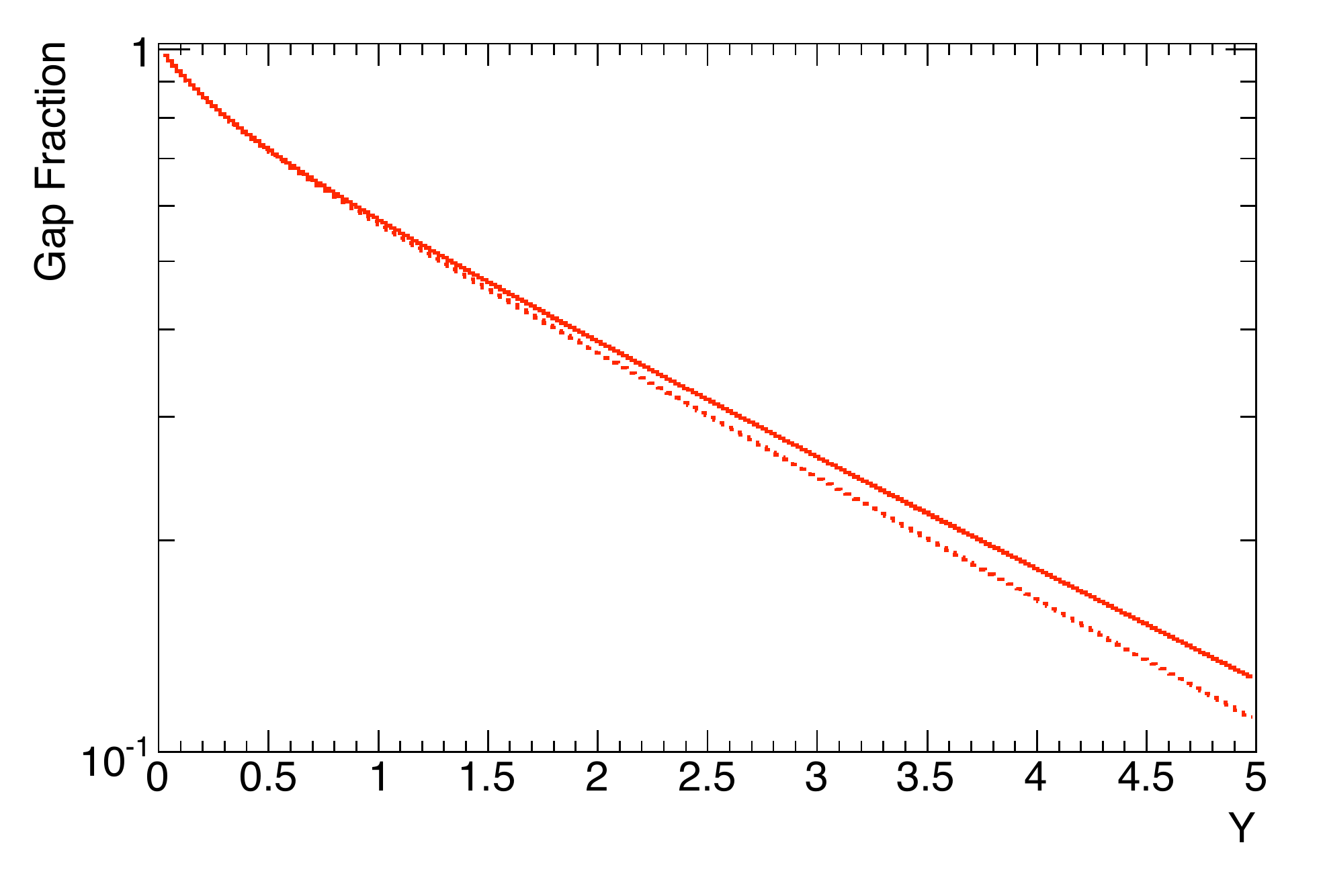}
\includegraphics[width=0.49\textwidth, clip]{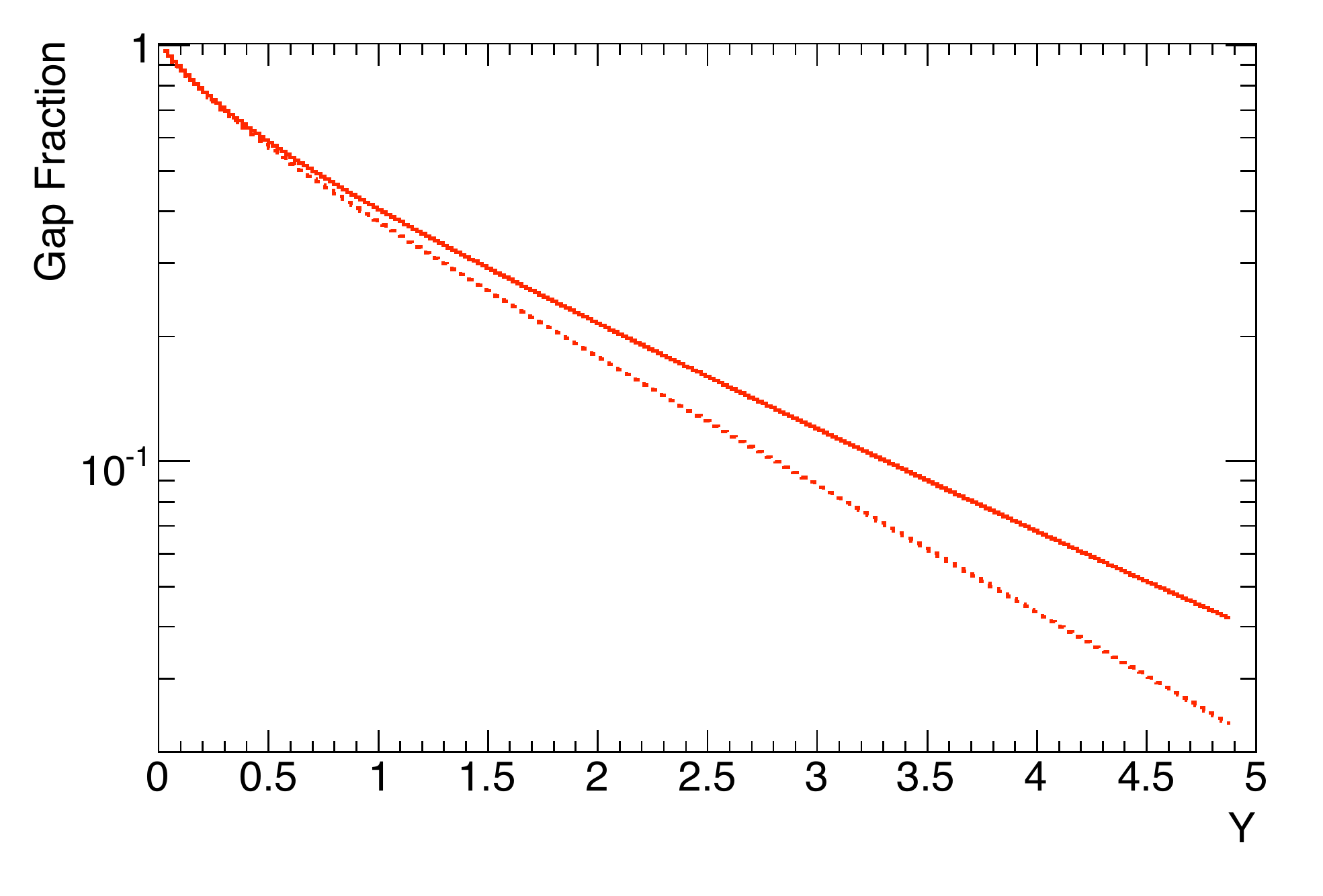}
\caption{Plot of the gap fraction as a function of $Y$, for $Q=100$~GeV (left) and for $Q=500$~GeV (right). The dashed lines are obtained by dropping the imaginary part of the evolution matrices.} \label{fig:kfact2}
\end{center}
\end{figure}

In Fig.~\ref{fig:kfact1} we show the gap cross-section, normalised to the Born cross-section (i.e. the gap fraction), as a function of $Q$ at two different values of $Y$ and in Fig.~\ref{fig:kfact2} it is presented as a function of $Y$ at two different values of $Q$. The solid lines represent the results of the (zero out-of-gap gluons) resummation; the dashed lines are obtained by omitting the $i \pi$ terms in the soft anomalous dimension matrices. Neglecting these contributions corresponds to neglecting the contributions from Coulomb gluons. As a consequence, the gap fraction is reduced by $7\%$ at $Q=100$~GeV and $Y=3$ and by as much as $50\%$ at $Q=500$~GeV and $Y=5$. Large corrections from this source herald the breakdown of the parton shower approach. This is as expected, for it is these terms that are ultimately reponsible for the colour mixing that generates colour singlet exchange, which dominates for large enough $Y$ and small enough $Q_0$.

\begin{figure} [th]
\begin{center}
\includegraphics[width=0.69\textwidth, clip]{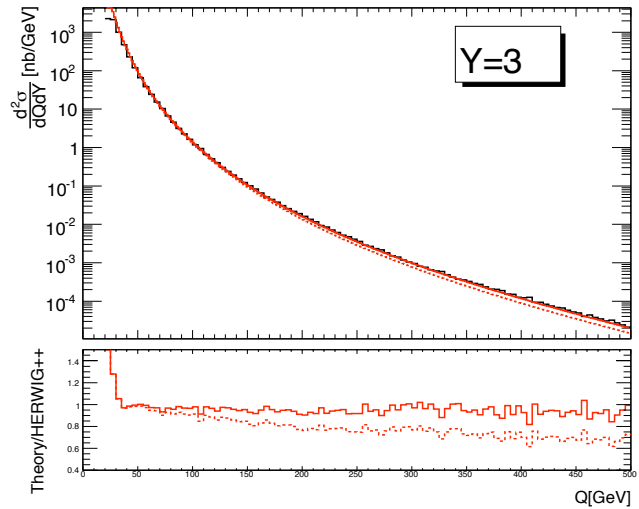}
\caption{Plot of the gap cross-section at $Y=3$. The resummed results (line types as in Fig.~\ref{fig:kfact1} and Fig.~\ref{fig:kfact2}) are compared to the result of \textsc{Herwig++}.} \label{fig:gapx}
\end{center}
\end{figure}

In Fig.~\ref{fig:gapx} we compare the gap cross-section obtained after resummation (the line types are as in Fig.~\ref{fig:kfact1} and Fig.~\ref{fig:kfact2}) to that obtained using \textsc{Herwig++} after parton showering ($Q$ is taken to be the mean $p_T$ of the two leading jets). The broad agreement is encouraging and indicates that effects such as energy conservation are not too disruptive to the resummed calculation. Nevertheless, the histogram ought to be compared to the dotted curve rather than the solid one, because \textsc{Herwig++} does not include the Coulomb gluon contributions. We do expect the imposition of energy conservation to enhance the resummed gap cross-section. In addition, the influence of non-global logarithms (some of which are included in the Monte Carlo) should act in the opposite direction and push down the dotted curve. Finally, there is the fact that the Monte Carlo operates only at leading $N_c$ to consider. As anticipated, there are very significant corrections to the parton shower Monte Carlos at the larger values of $Q/Q_0$. Of course the resummation would benefit from matching to a NLO calculation and this should be done before comparing to data.

\section{Super-leading logarithmic contributions} \label{sec:one}
\begin{figure}
\begin{center}
\includegraphics[width=0.49\textwidth, clip]{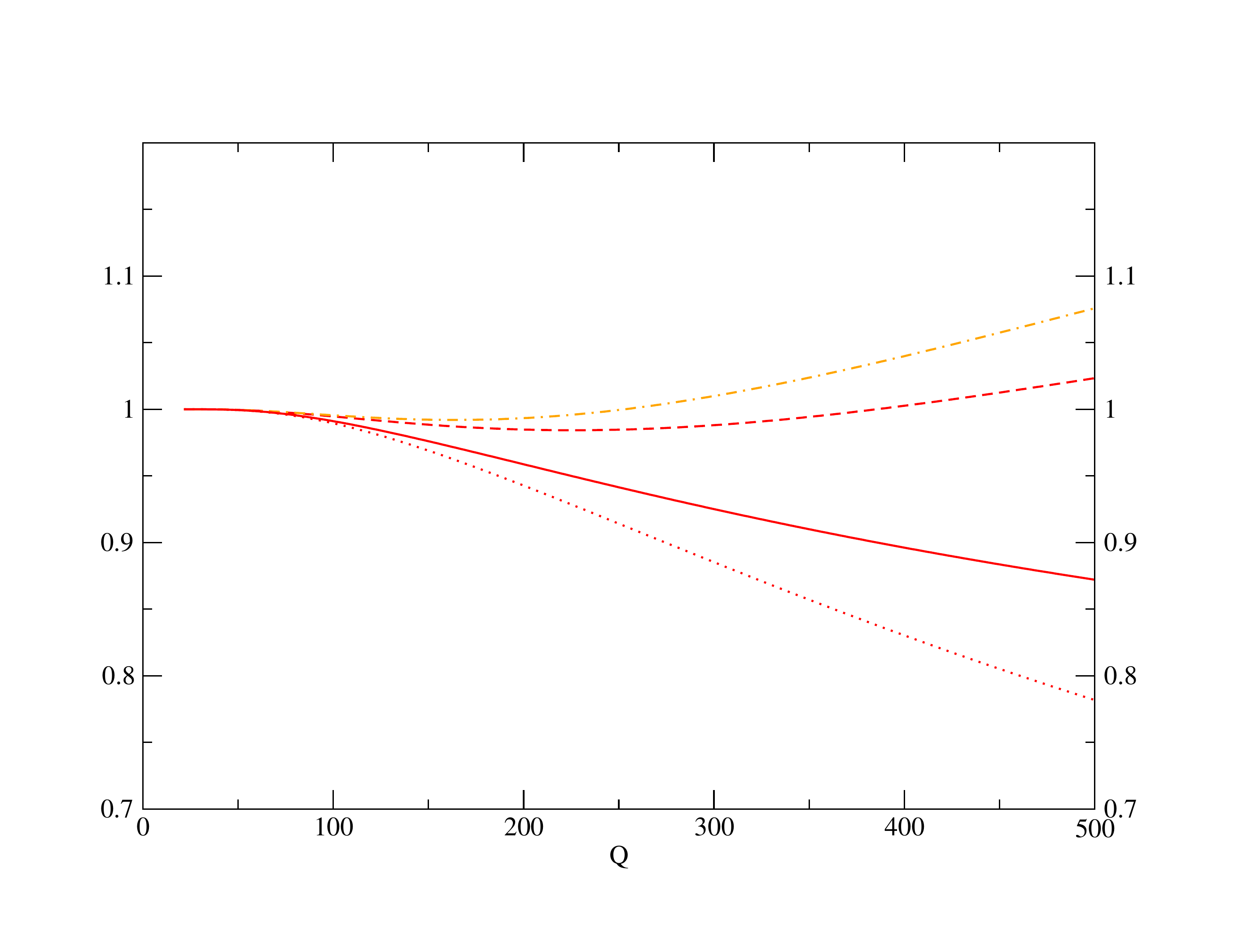}
\includegraphics[width=0.50\textwidth, clip]{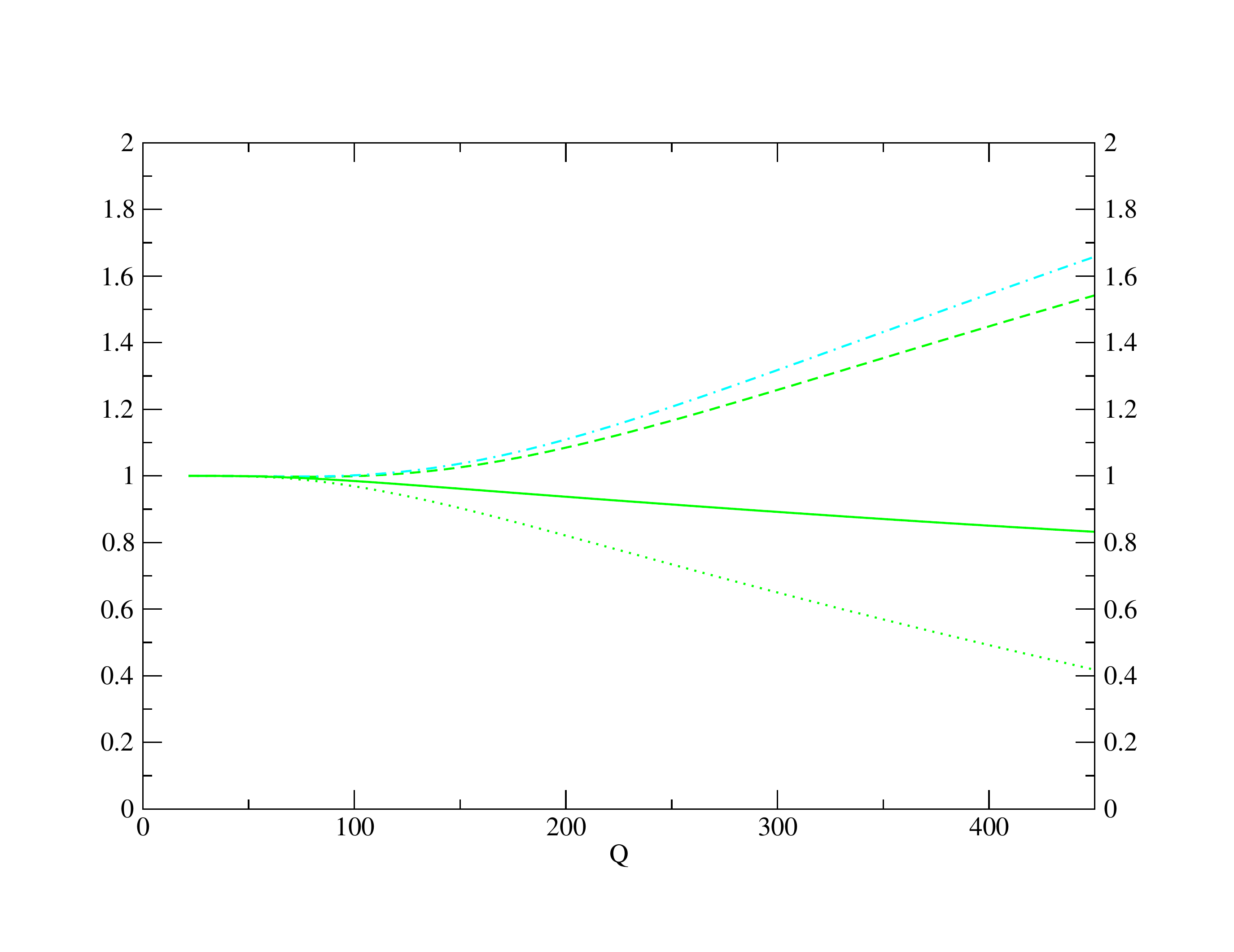}
\caption{Plots of the $K$-factors as a function of $Q$, for $Y=3$ on the left and for $Y=5$ on the right.  The curves are obtained by considering the one out-of-gap gluon cross-section at $\ord(\as^4)$ (dotted) and $\ord(\as^5)$ (dashed). The two-gluons outside of the gap contribution is included in the dash-dotted curves. The solid line corresponds to the resummation of the one out-of-gap gluon contributions and is explained in Section \ref{sec:resum}.} \label{fig:kfactfixedY}
\end{center}
\end{figure}

\begin{figure}
\begin{center}
\includegraphics[width=0.49\textwidth, clip]{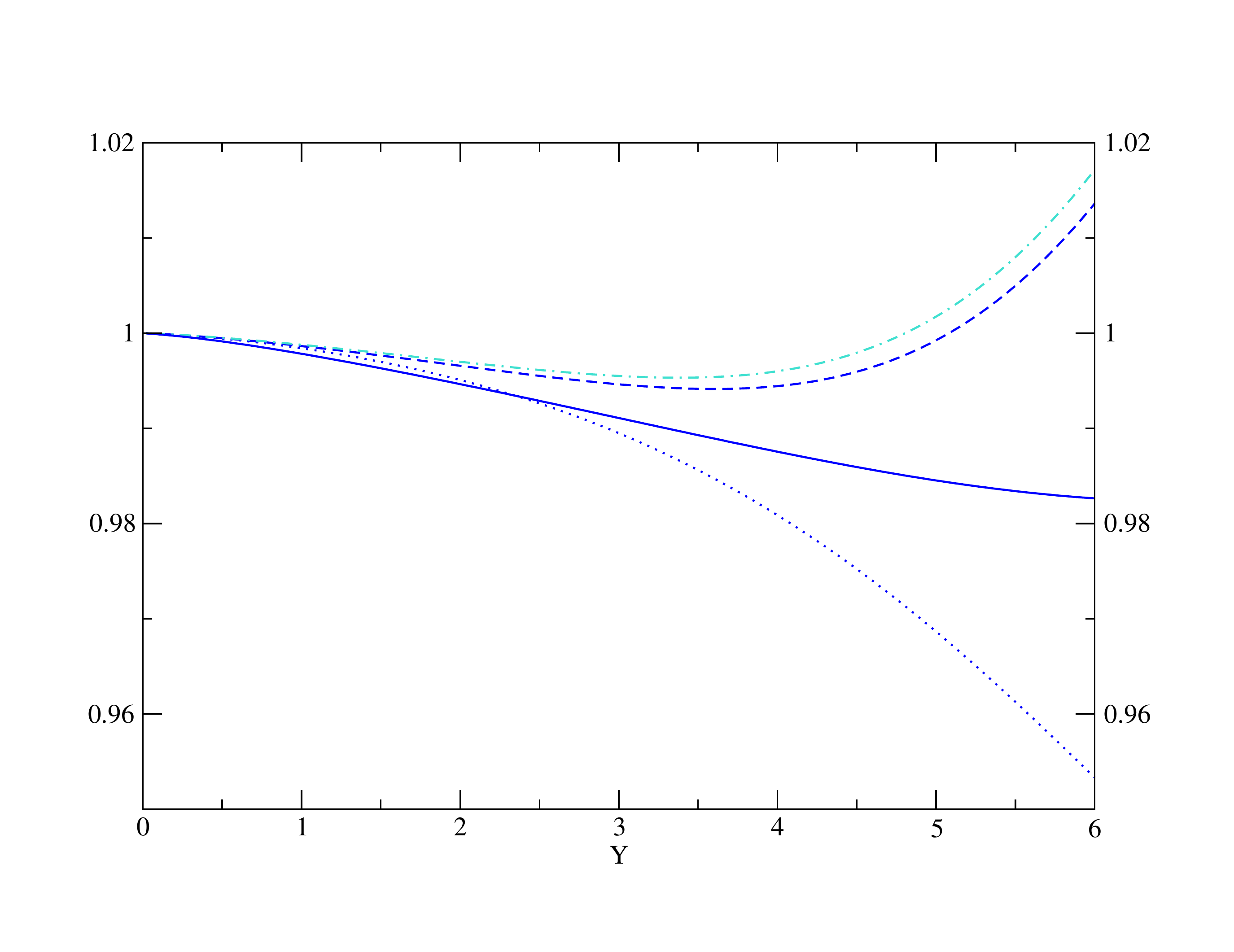}
\includegraphics[width=0.50\textwidth, clip]{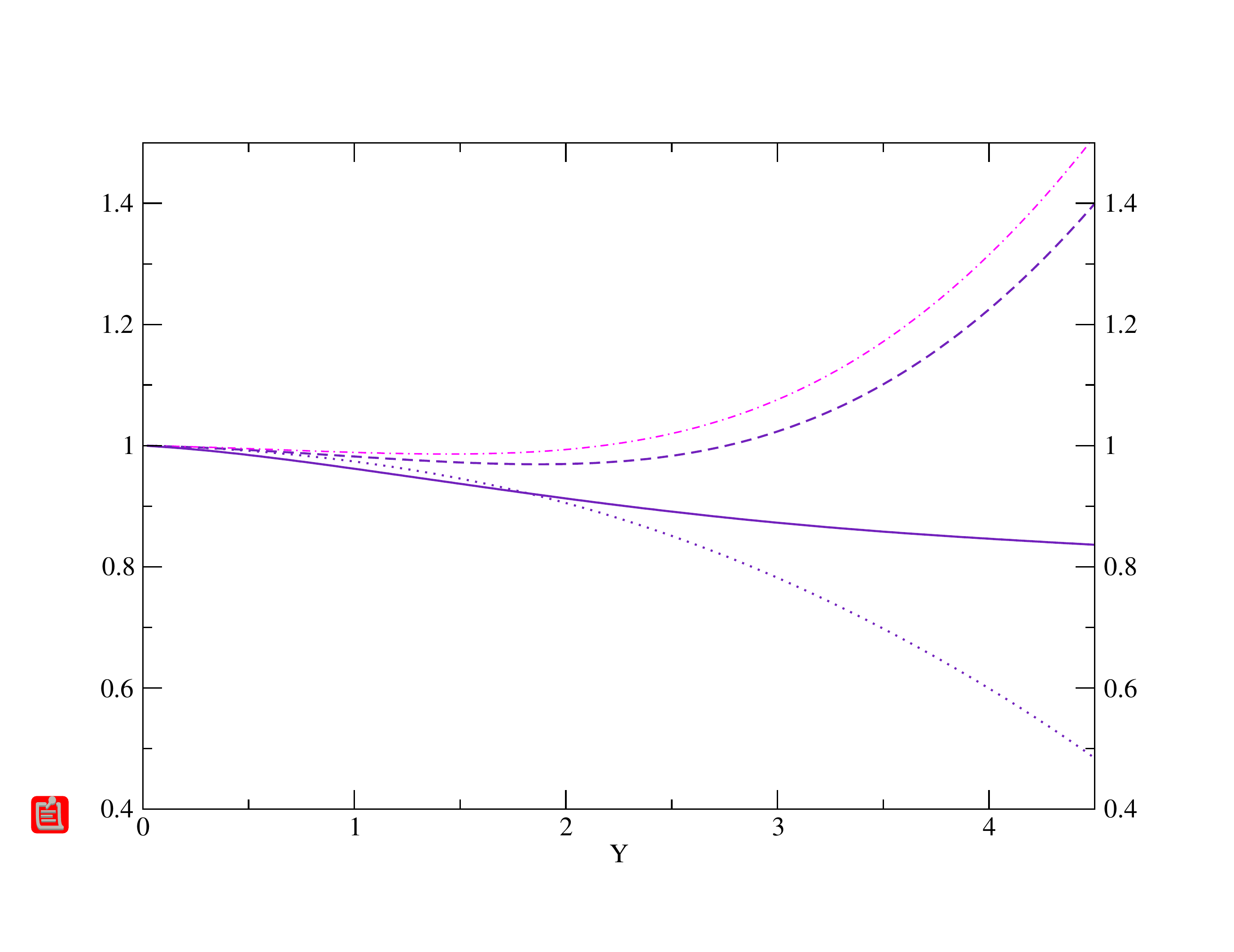}
\caption{Plots of the $K$-factors as a function of $Y$, for $Q=100$~GeV on the left and for $Q=500$~GeV on the right. The curves are obtained by considering the one out-of-gap gluon cross-section at $\ord(\as^4)$ (dotted) and $\ord(\as^5)$ (dashed). The two-gluons outside the gap contribution is included in the dash-dotted curves. The solid line corresponds to the resummation of the one out-of-gap gluon contributions and is explained in Section \ref{sec:resum}.}\label{fig:kfactfixedQ}
\end{center}
\end{figure}

In this section we shall evaluate the effect of including the super-leading logarithms (SLL). In~\cite{SLLfixed}, the coefficients of the SLL have been computed order-by-order in perturbation theory up to $\ord \left( \as^5\right) $, with respect to the Born cross-section.\footnote{More precisely, the calculation is performed in the high energy limit, considering only the $t$-channel gluon exchange contributions to the Born cross-section.} At this order in the perturbative expansion one must consider the contributions coming from one and two gluons outside the rapidity gap. We start by computing the impact of the super-leading contributions on the hadronic cross-section order-by-order in perturbation theory. To this end we define three different $K$-factors:
\bea \label{kfactfixed}
K^{(1)}_{\as^4}&=& \frac{\sigma^{(0)}+\sigma^{(1)}_{\as^4}  }{\sigma^{(0)}} \,,\nonumber \\
K^{(1)}_{\as^5}&=& \frac{\sigma^{(0)}+\sigma^{(1)}_{\as^4}+ \sigma^{(1)}_{\as^5}  }{\sigma^{(0)}} \,,\nonumber \\
K^{(2)}_{\as^5}&=& \frac{\sigma^{(0)}+\sigma^{(1)}_{\as^4}+\sigma^{(1)}_{\as^5}+\sigma^{(2)}_{\as^5}}{\sigma^{(0)}} \,.\nonumber \\
\eea
The SLL contributions to the cross-section are the ones computed in~\cite{SLLfixed}:
\bea\label{fixedSLL}
\sigma^{(1)}_{\as^4}    & =& -\sigma^{\rm born} \left(\frac{\as}{\pi} \right)^4 \frac{\mathcal{C}_1}{5 !} \pi^2 Y \left(\ln \frac{Q}{Q_0} \right)^5 \,, \\  
 \sigma^{(1)}_{\as^5}  & =&+   \sigma^{\rm born} \left(\frac{\as}{\pi} \right)^5 \frac{\mathcal{D}_1}{6!} \pi^2 Y^2  \left(\ln \frac{Q}{Q_0} \right)^6 \,,  \\  
\sigma^{(2)}_{\as^5}    & =& +\sigma^{\rm born}\left(\frac{\as}{\pi} \right)^5 \frac{\mathcal{C}_2}{7!} \pi^2 Y  \left(\ln \frac{Q}{Q_0} \right)^7 \,,
\eea
where $\mathcal{C}_i$ and $\mathcal{D}_i$ are colour factors.  Notice that we have chosen the notation in order to emphasize the factorial denominators coming from the strongly-ordered $k_T$-integrals. We plot these $K$-factors in Fig.~\ref{fig:kfactfixedY} as a function of $Q$, for $Y=3$ (left) and for $Y=5$ (right).  The curves on the left (right) are obtained by considering one out-of-gap gluon at $\ord(\as^4)$ (dotted) and $\ord(\as^5)$ (dashed). The two-gluons outside the gap contribution is included in the dash-dotted curves.
We also plot these $K$-factors for two values of $Q$ as a function of the rapidity separation $Y$ in Fig.~\ref{fig:kfactfixedQ}. The plot on the left is for $Q=100$~GeV, while the one on the right is for $Q=500$~GeV. The different line styles are as in Fig.~\ref{fig:kfactfixedY}. 
From these plots it is clear that, for jets with rapidity separation $Y \ge 3$ and transverse momentum bigger than $200$ GeV, the inclusion of fixed order SLL contributions leads to sizeable  but unstable $K$-factors. At every order in perturbation theory, for a given number of out-of-gap gluons, the SLL contribution acquires an extra power of $Y$ and $L$; the colour factor also increases. For large enough rapidities and transverse momenta, higher order contributions become important and their resummation is necessary. Currently the resummation of SLL contributions has been carried out only for the case of one gluon outside the gap, and we will discuss it in the next section.  Figures \ref{fig:kfactfixedY} and \ref{fig:kfactfixedQ} offer some hope that the impact of the two-or-more gluons outside the gap contribution may be modest, since the difference between the $K$-factors which include the two-out-of-gap gluon contributions (dash-dotted curves) and the ones that do not (dashed curves) is not large.

\subsection{Resummation of one gluon outside the gap contribution\label{sec:resum}}
The aim of this section is to resum the SLL contributions that arise as a result of allowing one soft gluon outside the rapidity gap. The general framework is described in~\cite{SLL1} and~\cite{SLLind}. One must now consider both real and virtual corrections to the four-parton scattering, each dressed with any number of soft (either Coulomb or eikonal) gluons:
\beq \label{master1}
|\m_1|^2 = - \frac{2 }{\pi} \int_{Q_0}^{Q} \frac{d k_T}{k_T}\as(k_T) \int_{\rm out} \left( \Omega_R + \Omega_V\right)\,.
\eeq
Super-leading logarithms originate when the out-of-gap gluon becomes collinear with one of the initial state partons. 

Since our aim is the resummation of the SLL, we can work in the limit where the gluon is collinear to one of the initial state partons. We also consider only leading terms in the rapidity separation $Y$, both in the hard scattering and anomalous dimension matrices. As a result, the expressions for real and virtual emission simplify considerably.

First we extract the logarithm that arises after integrating over the rapidity of the out-of-gap gluon, i.e.
\beq \label{master2}
|\m_1^{\rm SLL}|^2 = - \frac{2 }{\pi} \int_{Q_0}^{Q} \frac{d k_T}{k_T}\as(k_T) \left(  \ln \frac{Q}{k_T} \right)  \left( \Omega^{\rm coll}_R + \Omega^{\rm coll}_V\right)\,.
\eeq  
In the case that the out-of-gap gluon is virtual, we need the operators to insert the out-of-gap gluon ($\boldsymbol{\gamma}$) and to evolve the four-parton matrix element (i.e. $\mathbf{\Gamma}$ in Eq.~(\ref{gammaoperator})). In the collinear (and large $Y$) limit they are
\bea 
\boldsymbol{\gamma} &=& \frac{1}{2} \mathbf{t}^2_i \;, \quad i =1,2\,, \\
\mathbf{\Gamma} &=& \frac{1}{2} Y \mathbf{t}^2_t + i \pi \mathbf{t}_1\cdot \mathbf{t}_2\,.
\eea
The diagonal nature of $\boldsymbol{\gamma}$ ensures that the virtual contribution turns out to be the same as the four-parton case multiplied by a Casimir:
\beq
 \Omega^{\rm coll} _V=  \frac{1}{V_c} \langle m_0 |   \mathbf{t}^2_i e^{- \xi \mathbf{\Gamma}^{\dagger}}e^{- \xi \mathbf{\Gamma}} |m_0 \rangle\,.
\eeq
This result coincides with our expectation that a soft and collinear emission leads to an extra logarithm whose coefficient is the Casimir, multiplied by the lower order result. If the DGLAP plus-prescription would work, the real emission case should give the same contribution but with opposite sign. Interestingly, this does not occur. In particular, the operator which describes the emission of a real and collinear gluon, $\mathbf{D}^{\mu}_a$, and the five-parton evolution matrix, $\mathbf{\Lambda}$, are
\bea \label{5indep}
\mathbf{D}^{\mu}_a &=& (h_i-h)^{\mu} \mathbf{t}^a_i \;, \quad i =1,2\,,  \nonumber \\
\mathbf{\Lambda} &=& \frac{1}{2} Y \mathbf{T}^2_t + i \pi \mathbf{T}_1\cdot \mathbf{T}_2\,.
\eea
The cross-section depends only on $(h_i-h) \cdot (h_i-h)=-1$ and the real contribution is thus
\beq \label{omrind}
 \Omega^{\rm coll} _R= - \frac{1}{V_c} \langle m_0 | e^{- \xi(k_T,Q) \mathbf{\Gamma}^{\dagger}}
 {\mathbf{t}^a_i}^{\dagger}  e^{- \xi(Q_0,k_T) \mathbf{\Lambda}^{\dagger}} e^{- \xi(Q_0,k_T) \mathbf{\Lambda} }\mathbf{t}^a_i
 e^{- \xi(k_T,Q) \mathbf{\Gamma}} |m_0 \rangle\,.
\eeq
The non-commutativity of $\mathbf{T}_t^2$ and $i \pi \mathbf{T}_1 \cdot \mathbf{T}_2$ prevents us combining the exponentials and means that the contribution to the cross-section from real emissions is no longer proportional to a Casimir. At fourth order relative to the Born term, this triggers the breakdown of the plus-prescription and generates the super-leading logarithms.

The five-parton evolution matrices ($\mathbf{\Lambda}$) have been computed in~\cite{EVqqqqg} for $qq \to qqg $ and more recently in~\cite{5partons} for the other cases. We have repeated the calculation but do not show the explicit results here because the anomalous dimension matrices can be quite big, e.g. in the case of five gluons we have a $22 \times 22$ matrix~\cite{5partons}. Before presenting the results of the resummation, some comments are due regarding the operator 
$\mathbf{T}^2_t$. As in the four-parton case, it represents the colour exchanged in the $t$-channel and we use
\beq
\mathbf{T}^2_t=(\mathbf{T}_1 + \mathbf{T}_3 + \mathbf{T}_k)^2= (\mathbf{T}_2 + \mathbf{T}_4)^2\,,
\eeq
when the gluon, denoted by the index $k$, is collinear to parton $1$, while
\beq
\mathbf{T}^2_t=(\mathbf{T}_2 + \mathbf{T}_4 + \mathbf{T}_k)^2= (\mathbf{T}_1 + \mathbf{T}_3)^2\,,
\eeq
when the gluon is collinear to parton $2$. As previously discussed, in order to implement the 
$t \leftrightarrow u$ symmetrisation these should be replaced by
\beq
\mathbf{T}^2_t=(\mathbf{T}_1 + \mathbf{T}_4 + \mathbf{T}_k)^2= (\mathbf{T}_2 + \mathbf{T}_3)^2\,,
\eeq
and
\beq
\mathbf{T}^2_t=(\mathbf{T}_2 + \mathbf{T}_3 + \mathbf{T}_k)^2= (\mathbf{T}_1 + \mathbf{T}_4)^2\,.
\eeq
However, after swapping $t \leftrightarrow u$, the hard scattering matrices are exponentially suppressed in~$Y$ and we can neglect them.
Once a matrix representation for the different operators has been found, Eq.~(\ref{omrind}) becomes
\beq
 \Omega^{\rm coll} _R= - {\rm tr} \left[ H e^{- \xi(k_T,Q) \Gamma^{\dagger}}
 D^{\dagger}  e^{- \xi(Q_0,k_T) \Lambda^{\dagger}} e^{- \xi(Q_0,k_T)\Lambda} D  e^{- \xi(k_T,Q) \Gamma}  \right] \,.
\eeq
We checked our results by expanding the exponentials in powers of $\as$ and comparing them  with the fixed order results of \cite{SLLfixed} up to $\ord (\as^5)$. In~\cite{SLLind} it was noticed that the coefficient of the SLL at $\ord(\as^4)$ was the same regardless of whether the out-of-gap gluon is collinear to parton $1$ or $2$. We find that this is no longer true for higher orders if the initial state partons are different.

Once the real and virtual contributions have been computed, for all sub-processes, it is straightforward to obtain the hadronic cross-section. We quantify the effect of resummation via a $K$-factor, i.e.
\beq
K^{(1)}= \frac{\sigma^{(0)}+\sigma^{(1)}}{\sigma^{(0)}} \,.
\eeq
The results are added (as the solid lines) to the plots in Fig.~\ref{fig:kfactfixedY} and Fig.~\ref{fig:kfactfixedQ}.
Generally the effects of the SLL are modest, reaching as much as 15\% only for jets with \mbox{$Q > 500$~GeV} and rapidity separations $Y > 5$. Remember that we have fixed the value of the veto scale $Q_0=20$~GeV and that the impact will be more pronounced if the veto scale is lowered. Finally, we ought to explain a particular feature of the resummed results  in comparison to fixed order. In  Fig.~\ref{fig:kfactfixedY}, the resummed curves lie closer to the lower order results than to the $\ord\left(\as^5\right)$ results, even at low $Q$, and in Fig.~\ref{fig:kfactfixedQ}, for small values of $Y$, the resummed curves dip below the $\ord\left(\as^4\right)$ curves. This behaviour is completely explained by the running of the coupling. The fixed order results are as in Eq.~(\ref{fixedSLL}), with $\as=\as(Q)$, while in the resummed case the running coupling is evaluated at $\as=\as(k_T)$, i.e. it sits inside the transverse momentum integration. 

\section{Conclusions and Outlook} \label{sec:conclusions}
In this paper we have studied the effects of soft gluon resummation on the implementation of a jet veto in the region between the hardest pair of jets in dijet events at the LHC. The bulk of the leading logarithms arise as a result of vetoing primary wide angle soft gluon emissions in the inter-jet region and we compare the result of a direct resummation of these with the \textsc{Herwig++} parton shower Monte Carlo. The agreement is generally good, however there are significant contributions arising from the exchange of Coulomb gluons, especially at large $Q/Q_0$ and/or large $Y$. Such corrections are not implemented in the parton shower Monte Carlos. Moreover, there is a need to improve the resummed results by matching to the fixed order calculation at NLO. These observations will have an impact on jet vetoing in Higgs-plus-two-jet studies at the LHC.

We also attempted to quantify the size of the super-leading logarithms that occur because gluon emissions that are collinear to one of the incoming hard partons are forbidden from radiating back into the veto region. Although only a partial calculation can be performed, we did manage to sum the super-leading logarithms to all orders in the case of one collinear emission. We found that the effects are generally modest but can reach 15\% for values of $Q/Q_0$ and $Y$ that are well within the reach of the LHC.

Although our attention has focussed on the role and resummation of wide-angle soft gluons, there remains much to be done before one can claim to have properly calculated the cross-section for ``gaps between jets''. In particular, it remains to match the resummed results to fixed order calculations at NLO, to systematically include the non-global logarithmic corrections and to include the high-energy (BFKL) logarithms at large $Y$ (a start in this direction was made in 
\cite{gapsBFKL}). Moreover, as discussed in \cite{Reffect1,Reffect2}, the correct implementation of the jet algorithm even corrects the Sudakov exponentiation of the primary emissions, e.g. it has been shown that $k_t$-clustering introduces an $R^3$-dependent correction and similar effects are expected using other algorithms.

\section*{Acknowledgements}
We thank Mrinal Dasgupta, Andrew Pilkington and Mike Seymour for many interesting discussions. We also thank the UK's STFC for financial support.

\end{document}